\documentclass[3p,twocolumn,preprint]{elsarticle}

\usepackage{lineno}
\usepackage[hidelinks,breaklinks]{hyperref}

\modulolinenumbers[5]

\usepackage{float}
\usepackage{framed}
\usepackage[utf8]{inputenc}
\usepackage[T1]{fontenc}
\usepackage[strict]{changepage}
\usepackage{csquotes}
\usepackage{xcolor}
\usepackage[most]{tcolorbox}
\usepackage[inline,shortlabels]{enumitem}
\usepackage{fdsymbol}
\tcbuselibrary{theorems}

\journal{Journal of Systems and Software}

\definecolor{formalshade}{gray}{0.96}
\definecolor{darkblue}{RGB}{0,80,155}
\definecolor{green}{RGB}{177,201,31}

\definecolor{ultramarine}{RGB}{219,219,219}
\definecolor{ultramarine}{RGB}{9,32,96}

\newtcbtheorem[number within=section]{mydef}{Definition}%
{colback=formalshade,colframe=darkblue,fonttitle=\bfseries,arc=0pt,outer arc=0pt,boxrule=0mm,toprule=0.5mm,bottomrule=0.5mm,left=1mm,right=1mm,
titlerule=0mm,toptitle=0mm,bottomtitle=1mm,top=0mm}{thm}

\definecolor{linequote}{RGB}{224,215,188}
\definecolor{backquote}{RGB}{249,245,233}
\definecolor{lightgray}{RGB}{238, 238, 238}
\definecolor{midgray}{RGB}{140, 141, 143}

\newtcolorbox{myquote}{%
    enhanced, breakable, 
    size=fbox,
    frame hidden, boxrule=0pt,
    sharp corners,
    boxsep=6pt,
    colback=lightgray,
    borderline horizontal={0mm}{0mm}{midgray},
    borderline horizontal={1pt}{0mm}{midgray}
}

\newtcolorbox{mybox}{
    enhanced, breakable, 
    size=fbox,
    frame hidden, boxrule=0pt,
    sharp corners,
    boxsep=6pt,
    colback=formalshade,
    borderline west={1pt}{0pt}{darkblue}
}

\newcommand{\para}[1]{\emph{$\smallblacktriangleright$ #1}}
\newcommand{\revised}[1]{{#1}}

\begin{document}

\begin{frontmatter}

\title{Privacy Explanations -- A Means to End-User Trust}

\author[1,2]{Wasja Brunotte\corref{cor1}}
\ead{wasja.brunotte@inf.uni-hannover.de}

\author[1]{Alexander Specht}
\ead{alexander.specht@inf.uni-hannover.de}

\author[1]{Larissa Chazette}
\ead{larissa.chazette@inf.uni-hannover.de}

\author[1,2]{Kurt Schneider}
\ead{kurt.schneider@inf.uni-hannover.de}

\cortext[cor1]{Corresponding author}
\address[1]{Leibniz University Hannover, Software Engineering Group, Hannover, Germany}
\address[2]{Leibniz University Hannover, Cluster of Excellence PhoenixD, Hannover, Germany}

\begin{abstract}
Software systems are ubiquitous, and their use is ingrained in our everyday lives. They enable us to get in touch with people quickly and easily, support us in gathering information, and help us perform our daily tasks. In return, we provide these systems with a large amount of personal information, often unaware that this is jeopardizing our privacy. End users are typically unaware of what data is collected, for what purpose, who has access to it, and where and how it is stored. To address this issue, we looked into how explainability might help to tackle this problem. We created privacy explanations that aim to help to clarify to end users why and for what purposes specific data is required. We asked end users about privacy explanations in a survey and found that the majority of respondents (91.6 \%) are generally interested in receiving privacy explanations. Our findings reveal that privacy explanations can be an important step towards increasing trust in software systems and can increase the privacy awareness of end users. These findings are a significant step in developing privacy-aware systems and incorporating usable privacy features into them, assisting users in protecting their privacy.
\end{abstract}


\end{frontmatter}


\section{Introduction}
\label{sec:intro}
Personal data has long become a sort of virtual currency~\cite{Rana2015, Patil2019}. In 2017, a headline in \emph{The Economist} stated that \enquote{the world's most valuable resource is no longer oil, but data} \revised{\cite{Parkins2017, Wieringa2021}}. This has allowed a thriving new industry to emerge, known as \emph{data brokers}. This lucrative and fast-growing industry treats data as a commodity. We generate enormous amounts of data every second, leaving digital traces of our online selves behind~\cite{Klitou2014}. \revised{All this data is usually stored, merged, and evaluated~\cite{Wieringa2021, Schneier2015Goliath, Dinev2014}.} For instance, when driving a car, data about the speed and the strength with which the driver steps on the pedals is collected. This data can be used for routine diagnosis or for accountability purposes in the event of an accident. A simple photo contains information such as timestamp, location (GPS coordinates), camera information, and settings that are often collected, processed, and stored with the actual image data~\cite{Schneier2015Goliath}. 

\revised{Unfortunately, government agencies and private companies that dispose of our data do not always use appropriate mechanisms to prevent accidental or intentional privacy violations~\cite{Bowman2015}. The \enquote{power} hidden in data has led institutions around us to rightly conclude something rather obvious: that this data has enormous value.} 

\subsection{\revised{The Privacy Dilemma}}
\label{sec:privacydilemma}

Every time someone uses a software system, they consciously or unconsciously make a \emph{trade-off} between the benefits of using the system and the data \revised{they} provide during use. The reason for this data exchange (and thus the disclosure of personal information) is often motivated by access to personalized content, \enquote{free} information~\cite{TunMin2016}, discounts~\cite{BARNETTWHITE2004}, and loyalty programs~\cite{Earp2003}, as well as other economic incentives~\cite{Hann2002}. Yet, this information disclosure usually happens without explicit (informed) consent, although this data is related to and belongs to the end user~\cite{Janssen2020}.

The majority of websites (over 60\% in Europe) rely on (cookie) \emph{consent notices} to get visitors' consent to their data practices. However, because the implementations of these consent notices have substantial usability flaws, it is frequently unclear to the end user what data is being collected, stored, and for what purpose~\cite{Utz2019, Soe2020}.

Hence, a more responsible approach to personal data is needed, both from the regulatory side and from the companies themselves. According to Garcia-Rivadulla~\cite{Garcia2016}, companies should not perceive this as a threat. Rather, they should see it as an opportunity to innovate in the context of privacy and gain consumer trust. \revised{In a study, Cummings et al.~\cite{Cummings2021} discovered that if users are given more information about how their data will be used, they are more willing to provide this data.}

\subsection{\revised{Data Economy and Privacy Awareness}}
\label{sec:privacyawareness}

With all these factors in mind, it is crucial to find \revised{alternative solutions} to this problem. As Bowman et al.~\cite{Bowman2015} state, \enquote{engineers have a responsibility to the rest of society}.
Therefore, even if there is no \enquote{right} answer about one's right on the level of privacy or on how privacy could be most effectively protected, software engineers should actively tackle this challenge and find answers to the existing open questions surrounding this topic. 

Following this line of thought, the \enquote{principle of minimum asymmetry} should be considered~\cite{Jiang2002}. According to this principle, \enquote{a privacy-aware system should minimize the asymmetry of information between data owners\footnote{\label{datanote}Jiang et al.~\cite{Jiang2002} defined the terms data owner, data collector, and data user in their work, which we adopt here. \emph{Data owners} are the individuals whose data is being used or accessed (e.g., end users). \emph{Data collectors} are individuals or systems that collect information about data owners. \emph{Data users} are individuals or systems that use (process) this information.}, data collectors\footnotemark[\value{footnote}], and data users\footnotemark[\value{footnote}]}. This should be done by \enquote{\emph{decreasing} the flow of information from data owners to data collectors and users} and \enquote{\emph{increasing} the flow of information from data collectors and users back to data owners}~\cite{Jiang2002}. 

To address today's challenges regarding privacy (data economy and benefits), it is critical to develop systems that are \emph{privacy-aware} in design, incorporate \emph{usable} privacy features, and enter into a transparent dialogue with end users about data practices. 

\subsection{\revised{Privacy Explanations as a Solution}}
\label{sec:peintro}

So far, privacy policies are the primary channel to inform users about data practices of a service provider. However, privacy policies are \revised{insufficient} when it comes to informing users since they are too long, too vague, and the information required can often only be interpreted with legal background knowledge~\cite{BrunottePrivacyPlugin, Jensen2004, PPLong2, PPUnderstanding1, PPUnderstanding2}. \revised{Up to this moment, there is a lack of a user-centered solution to explain privacy-related aspects in an appropriate and understandable way.}

Explainability is a non-functional requirement (NFR) that is increasingly seen as a means to mitigate a system's lack of transparency and provide an understanding of a system's behavior among end users by giving explanations and disclosing information~\cite{Chazette2021, Koehl2019}. 
Thereupon, explanations might be a means to inform end users about a system's data practices. For instance, if an app needs access to a user's location, explanations can inform the user about the purpose \revised{and scope} of the data collection.

To contribute to the research of privacy in software engineering, we follow our research agenda~\cite{Brunotte2021}\revised{, by employing the concept of explainability to bridge the gap between the process of disclosing personal information and the lack of transparency involved in this process.} \revised{To this end,} we conducted an online survey to assess whether there is a need among end users to receive explanations of privacy aspects, how they perceive such explanations, and whether they have an influence on end users' trust toward a system. \revised{The results of our study show that respondents are interested in privacy explanations. They consider them beneficial, and state that privacy explanations might contribute to increasing trust in software.}

This paper is structured as follows: in the following \sectionautorefname~(\ref{sec:background}), we present background and related work. In \sectionautorefname~\ref{sec:research}, we present our research questions (RQs), and outline the chosen research design. In \sectionautorefname~\ref{sec:results}, we \revised{present} the findings of our survey, in \sectionautorefname~\ref{sec:discussion}, we discuss our results and propose a forecast about future work (\ref{sec:futurework}). In \sectionautorefname~\ref{sec:threats} we \revised{discuss} threats to validity. Finally, we conclude our paper in \sectionautorefname~\ref{sec:conclusion}.

\section{Background and Related Work}
\label{sec:background}

\begin{table*}[t]
\centering
\caption{\revised{Overview of authors' interpretations of privacy and its impact on individuals}}
\label{tab:privacy}
\begin{tabular}{p{0.75\linewidth}|p{0.15\linewidth}}

\multicolumn{1}{c|}{\textbf{\revised{Authors' Interpretations of Privacy and its Impact}}} & \multicolumn{1}{c}{\textbf{Reference}} \\ \hline \hline

Privacy is \enquote{a legal right} and \enquote{the right to be let alone} & \cite{TheRightToPrivacy} \\ \hline

\enquote{Privacy is structured by the the answer . . . to the questions \enquote{who are the persons you wish to exclude from having this knowledge?}}  & \cite{PrivacyAUsefulConcept} \\ \hline

\enquote{Privacy is the claim of individuals, groups, or institutions to determine for themselves when, how, and to what extent information about them is communicated to others} & \cite{PrivacyAndFreedom} \\ \hline


\enquote{The desire by each of us for physical space where we can be free of interruption, intrusion, embarrassment, or accountability and the attempt to control the time and manner of disclosures of personal information about ourselves} & \cite{smith2000ben} \\ \hline

\enquote{The right to privacy exists because democracy must impose limits on the extent of control and direction that the state exercises over the day-to-day conduct of individual lives} & \cite{RightOfPrivacy} \\ \hline

\enquote{A state in which persons may find themselves} & \cite{Velecky1978} \\ \hline

Influence of a person's well-being & \cite{PsychologicalAspectsOfPrivacy, PrivacyWellBeing, PrivacyMeaning, PrivacyBiological} \\  \hline

It is important for our mental and physical health & \cite{Petronio2002} \\ \hline

Privacy is important to grow personally and its autonomy leaves room in determining one's own path in life & \cite{PrivacyMeaning} \\ \hline

\revised{Privacy is important for the endurance of a strong society} & \revised{\cite{PsychologicalAspectsOfPrivacy}} \\ \hline

\revised{It acts like some sort of glue that binds individuals together in healthy relationships} & \revised{\cite{Braeunlich2021}} \\ \hline

\revised{A loss of privacy is not only unsettling but also represents an insult to a person’s dignity, independence, and integrity} & \revised{\cite{Bhave2020, Bloustein1964}}

\end{tabular}
\end{table*}

In this section, we define terms and provide background information that are necessary for the further understanding of this work. \revised{For this purpose, we will start discussing the different dimensions of \emph{the concept of privacy} (\ref{sec:privacy}). Second, we delineate privacy and \emph{online privacy} in \ref{sec:onlineprivacy}. In doing this, we want to gain a better understanding of what these concepts imply in terms of quality aspects (NFRs) for software systems.} 
\revised{In \ref{sec:privacyexplanation}, we define the term privacy explanation. In \ref{sec:trust-trustworthiness}, we define trust and trustworthiness. Although both terms are often used interchangeably, we would like to show why it makes sense from our point of view to distinguish the two terms from each other. We conclude this background section with related work in \ref{sec:rw}.}

\subsection{The Concept of Privacy}
\label{sec:privacy}
Privacy is a normative concept. It is not something new to our modern times but has existed for a very long time and is deeply rooted in sociological, philosophical, legal, political, and economic traditions~\cite{IsPrivacyPrivacy}. In the past, when humans were still hunters and gatherers, it was of crucial importance to know where and when there was ripe fruit or where the next water source was located. Sharing this information only with certain individuals of one's own community could potentially ensure a group's survival~\cite{Harari2015}. 

Even though scholars from different disciplines have investigated the concept of privacy from different perspectives, there is still no unified view or definition regarding this concept~\cite{PrivacyAspects, PrivacyMeaning, SelfProtectionOnlinePrivacy, PrivacyIsAConcern, PrivacyAndTheComputer, PerspectivesOnPrivacy}.

\revised{We provide an overview of authors' interpretations of privacy and its impact on individuals in \tablename~\ref{tab:privacy}.} All of these views and expressions have one thing in common: that privacy is about the control of information about personal matters and the creation of private spaces, whether physical or mental. To summarize, privacy enables us to enter a state in which we can withdraw from society, either physically, mentally, or both. This withdrawal is socially tolerated, important for our well-being, and essential for a healthy society.

We combine the above-mentioned concepts of the various authors and scholars who have studied the notion of privacy in a definition. We want to achieve two goals with this definition: \emph{(a)} compile a concise working definition that captures the many nuances of privacy and \emph{(b)} build a shared understanding of what privacy actually means.

\begin{mydef}[label=privacy-definition]{Privacy}{}
    \textbf{Privacy} is a right, a claim, and a state in which an individual independently sets their boundaries and determines what \emph{personal matters}\footnote{Personal matters refer to personal information, thoughts, feelings, and habits} they wish to share with or withhold from other individuals or society. These boundaries can be created through physical or psychological means. During this voluntary and temporary withdrawal, the individual is protected within these boundaries from unwanted intrusion (physically or mentally), embarrassment, judgment, discrimination, accountability, and societal norms and constraints.
\end{mydef}

Notwithstanding that privacy should be considered as a right of every individual~\cite{TheRightToPrivacy, McCloskey1980, De2012internet}, there must also be regulatory entities that monitor and defend the strict observance of this right by legal means.

\subsection{Online Privacy}
\label{sec:onlineprivacy}
\revised{Our management of personal information has required ongoing adaptation and revision due to the rapid advancement of information and communication technology.} For instance, printing technology simplified the reproduction and distribution of private information. The lines between personal and public life are blurring and shifting as a result of digitization and Internet use. We communicate using e-mail, messengers, and social media; seek for answers to private and sensitive questions using search engines. \revised{As a result, we leave a substantial quantity of digital evidence about our routines, opinions, and attitudes behind, making our privacy heavily dependent on how these systems are designed~\cite{BrunottePrivacyPlugin, Cavoukian2009}.}

In the offline world (the physical world), we are mostly in control of our own privacy and protected by the normative concept of privacy\revised{, backed by social norms and legal traditions.} For instance, if two friends are chatting in the town square, at most, bystanders can overhear portions of what they are saying. However, they can choose to stand further apart to ensure that their conversation is private. \revised{However, in the online world (the internet/cyberspace), this strategy is inapplicable and ineffectual. The amount of communication data saved here is significantly greater than a brief chat in the town square since it is stored in redundant ways~\cite{Klitou2014}.} To gain control over privacy in cyberspace, users must \emph{actively} take care of their privacy themselves. Users cannot rely on legal systems and cannot expect other users to comply with their social and cultural norms~\cite{SelfProtectionOnlinePrivacy}. 

\revised{Effective self-protection involves being conscious about and taking measures toward one's own privacy protection, as well as implies a certain technical knowledge about what is at stake, what is relevant to protect, and how to do it.} However, not everyone has the necessary knowledge to identify privacy issues, and actively protecting one's privacy is also thought to be difficult and time-consuming~\cite{Rudolph2018}.

\revised{In our definition, \emph{privacy} is an individual's ability to control their physical or mental presence and their right to mental or physical seclusion. In contrast, \emph{online privacy} is about an individual's control over their personal information in virtual space and their right to withhold this information. To this end, we define online privacy as:}

\begin{mydef}[label=online-privacy-definition]{Online Privacy}{}
    \revised{\textbf{Online Privacy} is a right, a claim, and a state in which an individual (data owner) sets their boundaries and determines (decide or control) what \emph{privacy aspect}\footnote{A privacy aspect refer to personal information, thoughts, and feelings. With respect to online privacy but especially data or information about a person. Examples: name, address, bank data, GPS location, etc.} they wish to share with data collectors and data users, by whom it may be accessed, and at what point in time this occurs.}
\end{mydef}

\subsection{Privacy Explanations}
\label{sec:privacyexplanation}
Concerning one's privacy in software, long privacy policies or short privacy notices are often the only available sources of information to end users where they can (possibly) find out what happens to their data. End users rarely read or understand privacy policies~\cite{BrunottePrivacyPlugin, Jensen2004, PPUnderstanding1}. As a result, users rarely benefit from privacy policies in terms of learning about their privacy. Therefore, users need a different form of clarification and transparency with respect to their online privacy.

Explainability is seen as an appropriate solution to mitigate the lack of transparency of a system~\cite{Chazette2019, jasanoff2017virtual, richardson2018survey}, it has an impact on the relationship of trust in a system and may lead to more end user acceptance~\cite{Chazette2021}. Privacy explanations can inform users what a system will or will not do with their personal data.

In this work, a privacy explanation does \textbf{\emph{not}} mean a privacy policy, privacy statement, or a privacy notice in the usual sense. We adopt the definition of explainable systems by Chazette et al.~\cite{Chazette2021} to define a \emph{privacy explanation} as follows:

\begin{mydef}[label=def-privacy-explanation]{Privacy Explanation}{}
    A \textbf{privacy explanation} is a corpus of information $I$ that a system $S$ gives to an addressee $A$ in context $C$ to explain the purpose $P$ for using a privacy aspect $X$.
\end{mydef}

The explanation $I$ is intended to provide an explanation to the end user (addressee $A$), i.e., a reason why the user's privacy-related information (privacy aspect $X$) is needed. This could be, for example, why a smartphone app needs the user's location. It is important that this explanation  provides a rationale (the purpose $P$) for why $X$ is needed and does so in a transparent and understandable way. This explanation might be expressed in text, graphics, audio, or any combination of those. The context is the situation in which an explanation is given, consisting \enquote{of the interaction between a person, a system, a task, and an environment}~\cite{Chazette2021}.

\subsection{Trust and Trustworthiness}
\label{sec:trust-trustworthiness}

Trust also plays an essential role in requirements engineering (RE), and system design~\cite{ISO25022, Giorgini2004, Golnaz2009}. Many see the concept of explainability as an appropriate means to amplify trust in a system, respectively stakeholder trust~\cite{Chazette2021, Chazette2020, LANGER2021}. Explainability has been identified as an NFR in the RE community~\cite{Chazette2021, Chazette2020, Koehl2019}, and it is often connected with trust in the literature due to its potential to increase trust~\cite{Nagulendra2016, Chakraborti19, Dahl2018, Floridi2018}. In light of this, it might be more appropriate to engineer and elicit requirements for explainability than to have requirements for trust directly. 

According to Kästner et al.~\cite{Kaestner2021} is \emph{trust} \enquote{an attitude a stakeholder holds \emph{towards} a system}. In contrast, the authors describe \emph{trustworthiness} as \enquote{a property of a system: intuitively, a system is trustworthy for a stakeholder when it is warranted for the stakeholder to put trust in the system}. In light of this, a system should \enquote{work properly} in a given context. Especially with regard to privacy, it is important to consider and differentiate trust and trustworthiness because if end users are to trust systems, they must also be sure that these systems are trustworthy.

\subsection{Related Work}
\label{sec:rw}

\revised{Houghten~\cite{Houghton2010} conducted a survey in which participants were asked to provide information on privacy concerns regarding social networks. The participants' biggest concern was the loss of control over their data, followed by the risk that supposed friends could turn out to be fraudsters.}

\revised{Anton et al.~\cite{Anton2010} asked participants in surveys about their privacy concerns when using software systems. The first survey began in 2002 and another survey was conducted in 2008. The result was that participants' concerns did not change, but the level of concern increased.}

\revised{Wilkowska et al.~\cite{Wilkowska2020} conducted a survey with users of a daily living app and an application for people suffering from dementia. In their study, the subjects were asked, among other things, who they would trust with their data and whether the storage of the data was of interest. The result was that doctors and family members were perceived as trustworthy. When it comes to storing data, information about where and for how long the data is stored was considered essential. The storage of data in the cloud was disagreed with by the majority of participants.}

\revised{The study by Wirth et al.~\cite{Wirth21} examines the extent to which individual laziness affects privacy. The results showed that individuals who are predominantly lazy are more likely to disclose their data as well as not change their privacy settings, despite the fact that changes were made to data practices. Lazy people are also more likely to share their data. According to the study, however, hardworking people are more likely to be responsible with their data.}

\revised{In the domain of artificial intelligence (AI), explainability has long played an important role (eXplainable Artificial Intelligence, XAI).}

\revised{In~\cite{Haroon2021} and~\cite{Smart2021}, approaches are being pursued to improve people's understanding of data protection. AI systems are used for this purpose. Haroon et al.~\cite{Haroon2021} focus lie on elderly persons to prevent a cognitive overload while using different ambient assisted living systems. Smart et al.~\cite{Smart2021} provide an overview of different strategies to avoid privacy threats.}

\revised{The works~\cite{Enyan2022} and~\cite{Binzhe2022} deal with privacy, fairness and explainability in the context of graph neural networks (GNNs). The authors mention in~\cite{Enyan2022} a positive interaction between explainability and privacy and that there is also a connection with trust. However,~\cite{Binzhe2022} refers to possible problems in terms of privacy, when providing explanations for whitebox GNNs, since the model parameters are accessible.}

\revised{A huge amount of research in the domain of XAI considers aspects such as fairness, trust, and privacy~\cite{BarredoArrieta2020, Amit2021, Balkir2022, Amparore2021, Tjoa2021, Mehdiyev2021}. Here, as in the studies mentioned above, the focus was less on the end user and how to offer them information about data practices in a meaningful way.}

\revised{In 2019, a study on the topic of explainability was conducted by Chazette et al.~\cite{Chazette2019}. Their work examined the influence of explanations in a navigation application. Chazette et al. come to the conclusion that explanations can significantly increase usability, but unnecessary information can have a negative effect.}

\revised{Cummings et al.~\cite{Cummings2021} used a survey to investigate whether users are more willing to share data if the disclosure of certain information is protected by differential privacy techniques. Their results showed that descriptions related to data use had a positive impact on respondents' willingness to share their data.}

\revised{Furthermore, a large body of research was done in terms of privacy policies and how to communicate them in a more comprehensible way to end users~\cite{Brunotte2022, Earp2005, Jensen2004, McDonald2009, Usmann2020, Keymanesh2021, Cheng2019, Zaeem2020}.}

\revised{With our study, we would like to contribute to closing a research gap regarding the information asymmetry between end users and software systems. Our goal is to investigate to what extent the concept of explainability can be employed to inform end users in a simple, comprehensible, and satisfying way how their personal data is used.}

\section{Research Goal and Design}
\label{sec:research}
Our research goal was to examine the perception of end users with respect to privacy explanations. In particular, we focused on the influence of such privacy explanations, whether they may foster end user trust toward a system and whether they may play a role in increasing end users' privacy awareness. To this end, we formulate the goal of our research according to the goal definition template by Wohlin et al.~\cite{Wohlin2012}.

\begin{mybox}
    \textbf{Goal definition:} We \emph{analyze} end users' \revised{perceptions} about the need for privacy explanations in software systems \emph{for the purpose of} investigating whether explanations might influence the level of trust \emph{from the point of view of} end users \emph{in the context of} an online questionnaire.
\end{mybox}

Based on \revised{our} goal definition, we framed our study into the following research questions (RQs). 

\begin{mybox}
	\textbf{RQ1:} How concerned are end users about their privacy and what threats regarding privacy are they worried about?\newline
	\newline \textbf{RQ2:} How do end users perceive explanations with respect to privacy aspects?\newline
	\newline \textbf{RQ3:} How are explanations of privacy aspects related to the concept of trust in a software system?
\end{mybox}

Online privacy is a significant issue because so much of modern life takes place online. As a result, we should concentrate on the creation of end user-centered privacy-aware systems. Accordingly, \textbf{RQ1} focuses on concerns and worries of end users with respect to their privacy. With this question, we wanted to get a general picture of what privacy risks users currently feel exposed to. \revised{Furthermore, this data should provide insight into whether end users' issues, or some of their concerns, can be mitigated by privacy explanations.}

\textbf{RQ2} focuses on how end users perceive privacy explanations, e.g., when an app communicates to its users why certain sensor data (location, etc.) from a smartphone is needed. \revised{On the one hand, we wanted to know whether users see the need for explanations when an app or service requires some privacy-related information. On the other hand, we wanted to know how users perceive it when a system provides an explanation regarding the use of a privacy aspect.} To address this, we provided the following hypothetical scenario to survey participants:

\label{hypotheticalsituation}
\begin{mybox}
    \textbf{Hypothetical Situation:} You heard from an acquaintance about a new app that lets you plan sightseeing tours or day trips for cities around the world. You decide to download this app to your smartphone for your upcoming city trip. However, when you start the app for the first time, it asks whether your location can be used and also asks for your date of birth. You are not sure about the reason, as the app does not give you any further information about the usage of your data.
\end{mybox}

In a first step, we asked participants if they would generally be interested in an explanation of \emph{why} the app asks for the data. In a second step, we gave the participants an exemplary explanation of the privacy aspects (regarding use of the location and date of birth). We then asked whether the explanations were perceived as useful. Finally, we asked the participants what information a privacy explanation should contain.

With \textbf{RQ3}, we want to find out how privacy explanations and end user trust are related. Therefore, we asked the participants what are the benefits of privacy explanations. These findings might enable us to understand the needs and expectations of privacy explanations and how to meet them. Furthermore, we wanted to investigate whether privacy explanations are a suitable means of providing transparency regarding data practices, since both understanding and transparency are quality aspects that can foster user trust in a system~\cite{Chazette2021, Koehl2019, ehsan2019automated}.

\begin{figure}[htbp]
	\centering
	\includegraphics[width=0.4\textwidth]{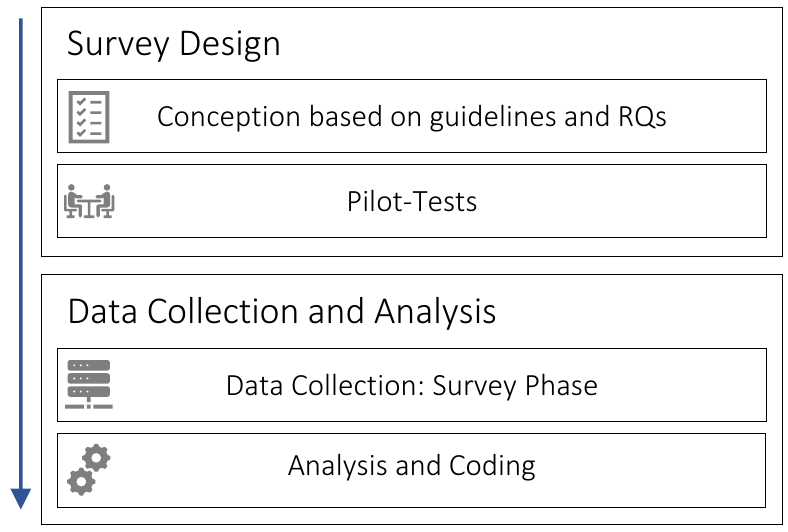}
	\caption{Overview of the research design}
	\label{fig:researchprocess}
\end{figure}

In order to answer the RQs, we structured our research design as shown in \figurename~\ref{fig:researchprocess}. Each phase will be described in the following sub sections.

\subsection{Survey Design}
\label{sec:survey-design}
To ensure the quality of our survey, we followed established guidelines for survey design~\cite{SudmanSurveyDesignGuidelines1982, JacobSurveyDesignGuidelines2014, Groves2011survey}. \revised{In order to avoid response bias, we took several actions in accordance with the guidelines. We tried to avoid leading questions, used precise and simple language, kept our questions short and clear as well as tried to use balanced and equal response ca\-te\-go\-ries. Furthermore, we informed the participants (in written form) that the survey is anonymous and they should answer honestly, as there are no right or wrong answers.}

\begin{figure}[htbp]
	\centering
	\includegraphics[width=0.4\textwidth]{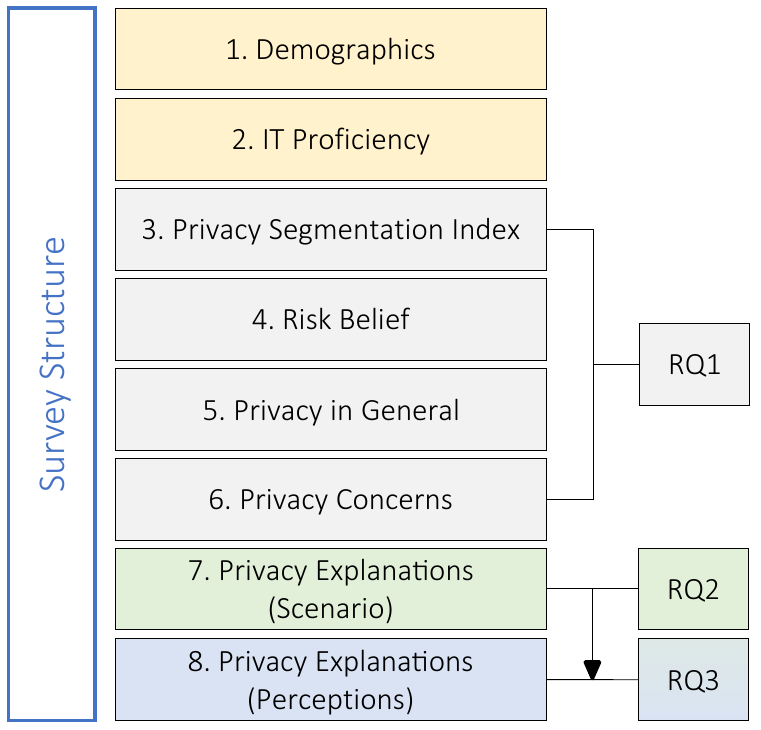}
	\caption{Overview of the survey structure}
	\label{fig:surveystructure}
\end{figure}

We defined the survey structure in line with our RQs. The survey started with a brief introduction and contained eight main parts with a total of 34 questions (30 multiple choice, four open-ended). Some of the multiple choice questions were also given the option for respondents to formulate their own answer text if none of the given answer options appealed to them. The structure of the survey, including the parts that answer the respective research questions, is depicted in \figurename~\ref{fig:surveystructure}.

The purpose of the first part (demographic questions) was to help to identify demographic factors that might have an impact on the respondents' answers. The second part contained questions to assess the respondents' experience with information technology (IT). The third part contained questions to collect respondents' \revised{Privacy Segmentation Index (PSI, Section~\ref{sec:psi-background})}. In the fourth part, we asked questions to help to assess participants' risk belief \revised{(Section~\ref{sec:rb})}. The fifth part comprised general questions about participants' privacy behavior. In the sixth part, participants were asked to describe situations in which they had concerns about their privacy when using software. In the seventh part of the survey, we gave the subjects a hypothetical scenario along with privacy explanations and asked questions regarding their perceptions. In the last part, we asked participants in general about possible benefits of privacy explanations as well as when a system should give such an explanation.

We conducted four rounds of pilot testing to assess the survey's quality. Two of these took place with members of our research group and two with candidates of the target population. Based on these pilots, we applied some minor corrections (e.g., addition of information regarding the interpretation of certain questions, minor text changes for better understanding).

\subsection{Data Collection and Analysis}
\label{sec:dca}
The data was collected via a web-based questionnaire. The survey was created with the survey tool LimeSurvey and hosted on our university's servers.

\subsubsection{Data Collection}
\label{sec:dc}
Data collection took place over two months, starting in June 2021. We distributed the survey through many means, including academic mailing lists, Facebook, and Twitter, and we invited our personal network to share the survey with their networks. Our target group was adult end users with different occupations and IT knowledge since we wanted to understand the perception of end users with different backgrounds on this topic. Because of our sampling strategy (contact networks mostly concentrated in Germany, Brazil, and abroad), we provided our survey in three languages: English, German and Portuguese. We expected that a large part of the participants would come from Brazil and Germany.

\subsubsection{Analysis and Coding}
\label{sec:analysis}
We applied qualitative and quantitative analysis techniques to the survey results. We exported the results to spreadsheets in order to calculate descriptive statistics. For the open-ended questions, we applied a qualitative data analysis consisting of an open coding approach, as described by Saldaña~\cite{Saldana2013}. \revised{According to Saldaña, coding is \enquote{one way of analyzing qualitative data} and it \enquote{transforms qualitative data into quantitative data, but it does not affect its
subjectivity or objectivity}~\cite{Seaman99}.}

We applied two consecutive coding cycles. First, we used \emph{In Vivo Coding}, it is a method \enquote{to preserve participants' meanings of their views and actions in the coding itself}~\cite{Charmaz2006}. \revised{In vivo codes serve as symbolic markers for the speech and meaning of the statements made by the respondents.}  We identified the essential passages of text in answers given by respondents in relation to the questions asked. A single response could result in more than one code. This was dependent on the length of the response as well as its meaning.

In the second coding cycle, we used \emph{Pattern Coding}. In this type of coding, summaries are grouped into smaller sets, constructs, or themes~\cite{Miles1994}. For this purpose, categories were formed to reflect the meaning based on the codes. While forming these categories, we tried to preserve respondents' opinions and avoid over-interpretation. \revised{These categories can help to understand implicit meanings and actions. In addition, comparisons can also be made between the data and the categories resulting from the codes.}

The coding procedure was conducted independently by the first two authors of this work. In cases of discrepancies, we discussed the differences until we reached consensus. We used Cohen's Kappa statistics~\cite{CohensKappa} to assess the reliability of the coding procedure. The resulting value of $\kappa=0.87$ showed an almost perfect agreement~\cite{LandisMeasurement}.

\subsection{Privacy Segmentation Index And Risk Beliefs}
\label{sec:psi-rb}
In our survey, we classified participants based on the Privacy Segmentation Index (PSI) as well as determined their Risk Beliefs. In the following, we describe the meaning of PSI and Risk Beliefs and how we determined them.

\subsubsection{Privacy Segmentation Index}
\label{sec:psi-background}
Westin developed the PSI to classify consumers according to their privacy concerns~\cite{Kumaraguru2005}. Although the PSI is related to a consumer perspective, it is adopted in broader contexts~\cite{Woodruff2014, Consolvo2005, TunMin2016}. With this in mind, we have also collected the PSI to capture respondents attitude towards their privacy concerns. The following statements are included in the PSI where respondents express their level of agreement through a 7-point Likert scale (\emph{strongly disagree} to  \emph{strongly agree}):

\begin{enumerate}[label=\textcolor{darkblue}{P{\arabic*}.}]
    \item Consumers have lost all control over how personal information is collected and used by companies.
    \item Most businesses handle the personal information they collect about consumers in a proper and confidential way.
    \item Existing laws and organizational practices provide a reasonable level of protection for consumer privacy today.
\end{enumerate}

Based on the responses, participants can then be classified one of three categories: \emph{Privacy Fundamentalists}, \emph{Privacy Pragmatists}, and \emph{Privacy Unconcerned}. The representative descriptions of these categories are given in the 2002 Harris report~\cite{Harris2002} as follows:

\textbf{Privacy Fundamentalists:} This group sees privacy as an especially high value, rejects the claims of many organizations to need or be entitled to get personal information for their business or governmental programs, thinks  more individuals should simply refuse to give out information they are asked for, and favors enactment of strong federal and state laws to secure privacy rights and control organizational discretion.

\begin{figure*}[t!]
	\centering
	\includegraphics[width=0.98\textwidth]{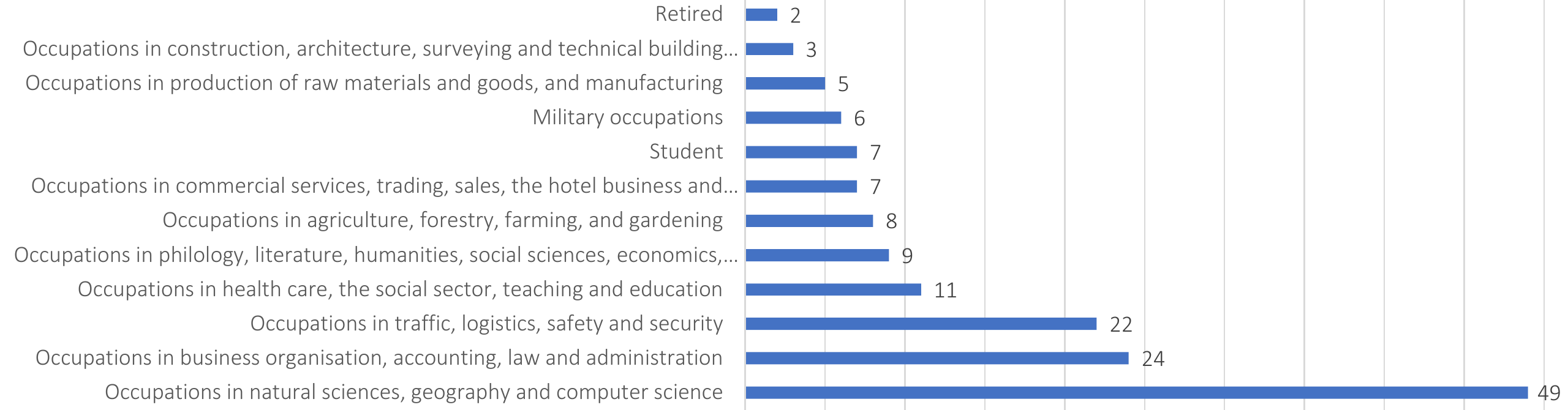}
	\caption{Overview of the occupations}
	\label{fig:country}
\end{figure*}

\textbf{Privacy Pragmatists:} This group weighs the value to them and society of various business or government programs calling for personal information, examines the relevance and social propriety of the information sought,  wants to know the potential risks to the privacy or security of their information, looks to see whether fair information practices are being widely enough observed, and then decides whether they will agree or disagree with specific information activities --- with their trust in the particular industry or company involved a critical decisional factor. The Pragmatists favor voluntary standards and consumer choice over legislation and government  enforcement. But they will back legislation when they think not enough is being done -- or meaningfully done -- by voluntary means.

\textbf{Privacy Unconcerned:} This group doesn't know what the \enquote{privacy fuss} is all about, supports the benefits of most organizational programs over warnings about privacy abuse, has little problem with supplying their  personal information to government authorities or businesses, and sees no need for creating another government bureaucracy (a \enquote{Federal Big Brother}) to protect someone's privacy.

According to Westin, the classification is as follows: Privacy Fundamentalists agree with P1 and disagree with both P2 and P3. Participants who disagree with P1 and agree with \revised{P2 and P3} belong to the Privacy Unconcerned class. The remaining participants can be classified as Privacy Pragmatists.

\subsubsection{Risk Beliefs}
\label{sec:rb}
Risk belief is a metric that can be used to quantify the level of risk a person perceives by sharing their information online. Tsai et al.~\cite{Tsai2006} used this metric in their work. We adopted their approach to calculate a \textbf{risk score} for each subject in the survey. Our aim has been to use the risk belief to possibly make a better assessment of the individual participants. For this purpose, we modified the questions of the risk beliefs a bit and formulated the questions not only in relation to online shopping, but in relation to online services in general. We asked respondents a total of four closed-ended questions for this purpose.
\begin{enumerate}[label=\textcolor{darkblue}{Q{\arabic*}.}]
    \item I feel safe giving my personal information to online services (such as online stores) and/or apps.*
    \item Providing online services or apps with personal information causes too many concerns.
    \item I generally trust online companies with handling my personal information, e.g., my purchase history.*
    \item How concerned are you about threats to your personal privacy online today?
\end{enumerate}
For questions one through three, we used a 7-point Likert scale to determine the level of agreement (\emph{strongly disagree} to \emph{strongly agree}). For the fourth question, we measured the level of concern with a 5-point Likert scale (\emph{not at all concerned} to \emph{extremely concerned}). Responses were scored according to the scales, with scores inverted for questions one and three to reflect feelings of concern (the higher the score, the higher the perceived risk). In order to map the 5-point Likert scale to the 7-point Likert scale, we weighted the items by 1.5. We see this mapping as justified and the participants' statements as not distorted. A chronbach's $\alpha$~\cite{Cronbach1951} value of 0.76 confirmed the reliability of the 7-item scale~\cite{George2009}.

\section{Results}
\label{sec:results}
In this section, we report the results related to our three RQs. In the \revised{first  subsection, we start by presenting the participants' demographics.} The subsequent subsections are devoted to each RQ, presenting the related results, and answering each RQ. 

\subsection{Demographics}
\label{sec:demographics}
We received a total of 209 responses. From the 209 participants who responded our survey, 155 completed the survey. For our data analysis, we considered only the 155 valid (complete) responses. Most respondents come from Germany (67.1\%) and Brazil (21.9\%). Ages ranged from 19 to 92 (M=39, SD=14.1). 

61.9\% of the participants identified themselves as male and 37.5\% as female, and one (0.6\%) to another gender. The majority of the participants work in the field of \enquote{natural science, geography and computer science} (32.03\%) as shown in \figurename~\ref{fig:country}. Two respondents did not answer this question. 

\para{IT proficiency.} In order to assess respondents' IT proficiency, we included self-assessment questions where the respondents had to indicate whether they are able to perform certain tasks and whether they are familiar with certain IT terms.

\begin{figure}[]
	\centering
	\includegraphics[width=0.47\textwidth]{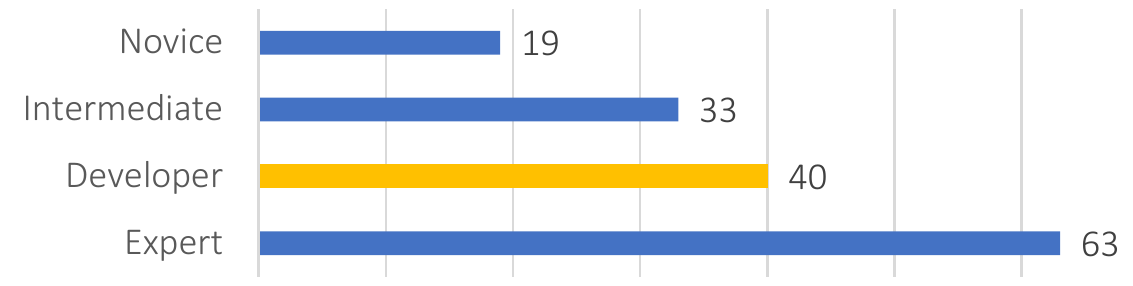}
	\caption{IT skill levels of the respondents}
	\label{fig:itskills}
\end{figure}

As shown in \figurename~\ref{fig:itskills}, the majority of the survey respondents (66.5\%, developers and IT experts) claimed to have a high level of IT proficiency. \revised{We have distinguished between \emph{developers} and \emph{experts}} because developers not only have profound IT knowledge but they also have sound knowledge about the internals and other programming aspects of software systems.

The results show that, in general, respondents are comfortable working with software systems. We grouped respondents into the group \emph{intermediate} if they claim to have at least significant software skills (such as creating functional spreadsheets or being able to quickly learn new programs) as well as knowledge of basic computing concepts. We assigned respondents who selected only the statement \enquote{I don't have that much experience, but I can check my email and do simple tasks with word processing software} as an answer to the \emph{novice} group.

\begin{figure}[]
	\centering
	\includegraphics[width=0.45\textwidth]{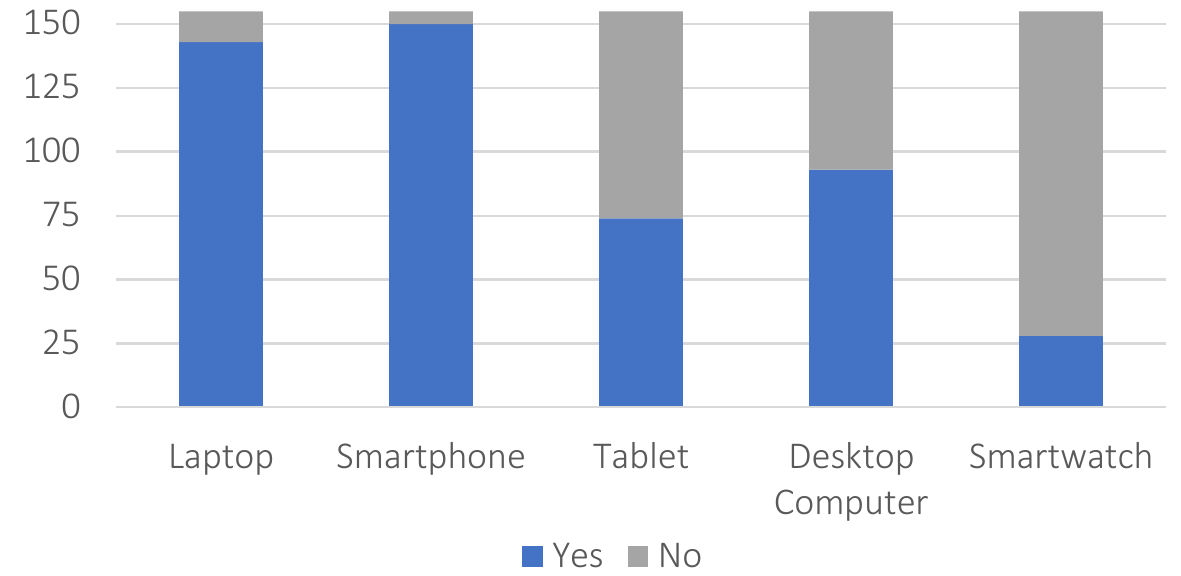}
	\caption{Usage of different devices}
	\label{fig:devices}
\end{figure}

\para{Device and software usage.} When asked about the usage of different devices (see \figurename~\ref{fig:devices}), 96.8\% of the respondents affirmed that they use a smartphone in their daily lives and all respondents use either a laptop or a desktop computer on a daily basis. 

\begin{figure}[]
	\centering
	\includegraphics[width=0.45\textwidth]{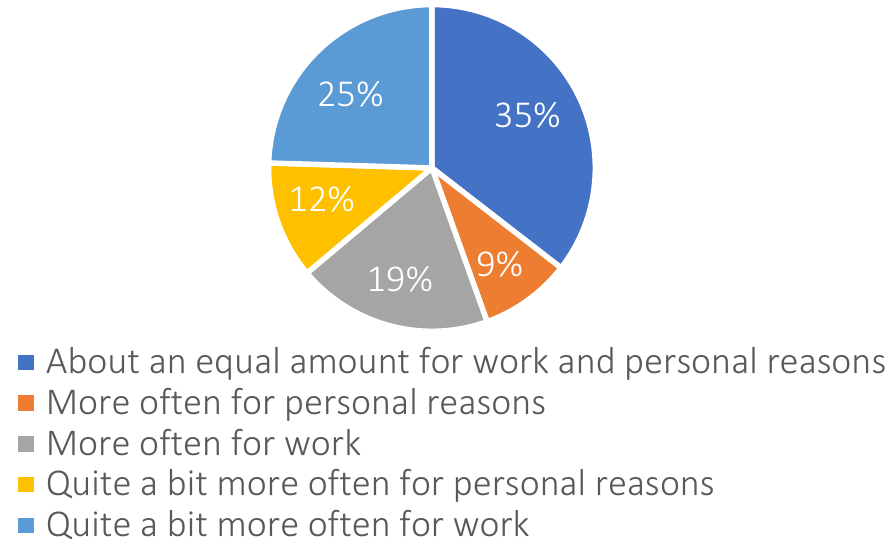}
	\caption{Usage of Software}
	\label{fig:softwareusage}
\end{figure}

When asked whether they use software systems more for work or personal reasons, 35.5\% of the respondents stated  that they use them more or less equally for work and personal reasons. 43.9\% use software more for work (19.4\% more often for work, 24.5\% quite a bit more often for work) as shown in \figurename~\ref{fig:softwareusage}.

Both factors (usage of different devices and usage of software) evidence how software systems play an integral part in the everyday lives of the respondents.

\subsection{RQ1 - Privacy Concerns}
\label{sec:rq1}
To determine respondents' concerns and worries respect to their privacy, we categorized respondents according to the PSI items, as described in Section ~\ref{sec:psi-background}. Subsequently, we quantified the level of risk our respondents perceive when sharing their information online (see risk beliefs, \ref{sec:rb}). Finally, we wanted to know what concerns the participants have regarding their privacy and what risks they see themselves exposed to. 

\subsubsection{Privacy Segmentation Index}
\label{sec:psi}

\begin{figure}[H]
	\centering
	\includegraphics[width=0.44\textwidth]{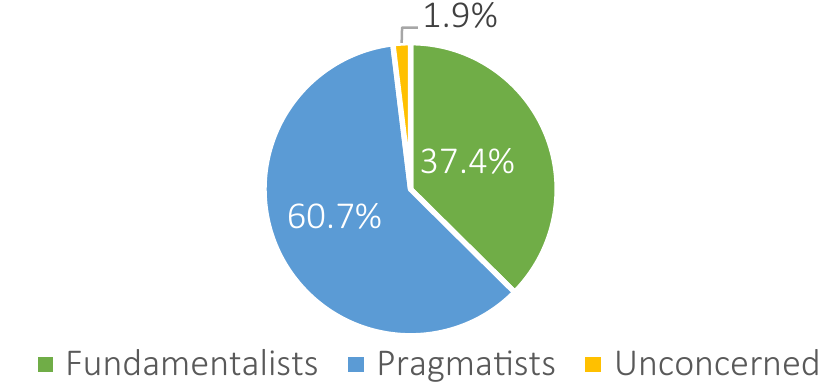}
	\caption{PSI of respondents}
	\label{fig:psi}
\end{figure}
\revised{\figurename~\ref{fig:psi} shows the distribution of the respondents according to the PSI items.}

\begin{figure*}[t!]
	\centering
	\includegraphics[width=0.8\textwidth]{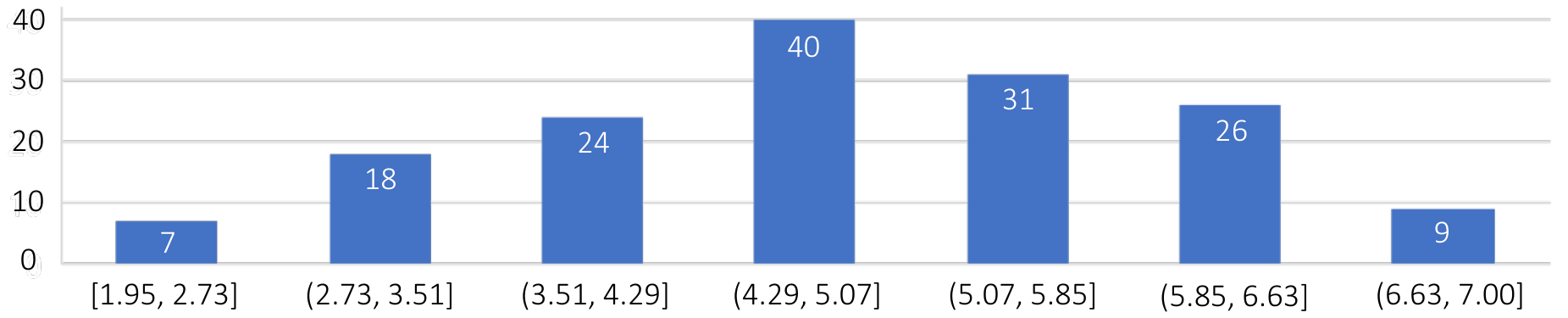}
	\caption{Histogram of risk scores}
	\label{fig:riskscores}
\end{figure*}

The distribution is largely consistent with data from Westin's privacy surveys (1996, 2000, 2001, and 2003)~\cite{Kumaraguru2005}, where the majority were classified as Privacy Pragmatists. The proportions of Privacy Fundamentalists and Privacy Unconcerned differ somewhat in our survey results. They show a higher proportion of Privacy Fundamentalists and a very low proportion of Privacy Unconcerned. Whether the Privacy Unconcerned have now become Pragmatists or Fundamentalists cannot be said on the basis of our data. The higher proportion of Privacy Fundamentalists could possibly be explained by the fact that people potentially value their (online) privacy more than it was the case in the past. The geographical distribution can also play a major role, since Germans are knowingly concerned about their online privacy~\cite{Schomakers2019}. But here, too, it is not possible to prove this assumption only on the basis of our data. 

\para{Required Permissions and Installing Software.} Following the questions on the PSI, we asked respondents whether they pay attention to \enquote{required permissions} while installing apps. 60\% indicate that they always pay attention, 34.2\% do sometimes, and 5.8\% do not pay attention. When asked how fast they press the \enquote{agree} button to terms and conditions when first using software, 30.3\% of the respondents state that they press \emph{instantly} the button, 55.5\% \emph{within one minute}, and 14.2\% \emph{spend more that one minute} before pressing the button.

50.0\% of the Fundamentalists and 65.9\% of the Pragmatists \emph{always} pay attention to \enquote{required permissions}. However, 66.6\% (2 respondents) of the Privacy Unconcerned stated that they also always pay attention to them (the high percentage results from the fact that of the 155 respondents, a total of 3 were classified as Privacy Unconcerned). The proportion that \emph{sometimes} pays attention is higher among Fundamentalists (41.3\%) than among Pragmatists (29.8\%).

When asked how fast they press \enquote{Agree} when installing software, the percentage who do so \emph{within one minute$^{\smalllozenge}$} or \emph{spend more than one minute$^{\medwhitestar}$} is also somewhat higher among Pragmatists (57.5\%$^{\smalllozenge}$, 14.9\%$^{\medwhitestar}$) than among Fundamentalists (53.5\%$^{\smalllozenge}$, 13.8\%$^{\medwhitestar}$), Privacy Unconcerned (33.3\%$^{\smalllozenge}$, 0\%$^{\medwhitestar}$).

Despite the fact that there is no significant difference between the groups, the results suggest that users are aware of privacy risks with respect to required permissions of apps because the majority pays attention to what permissions an app requires. When installing software, the majority also does not \enquote{agree} instantly. Our survey does not ask whether the users read information about data privacy during this time, for example. However, when this data is analyzed in conjunction with the question concerning needed permissions, we might infer that the respondents are well aware of the privacy threats that software may pose.

\para{Privacy Policies.} Privacy policies are the primary channel through which service providers inform end-users about their data practices. 56.1\% of the respondents \emph{rarely} to \emph{never} pay attention to whether a website provides a privacy policy, 23.9\% \emph{often} to \emph{always} pay attention, and 20\% pay attention \emph{sometimes}. In fact, only 8.4\% of respondents actually read a privacy policy (\emph{often} to \emph{always}), 13.5\% said they \emph{sometimes} read privacy policies and 78.1\% \emph{rarely} to \emph{never}.

\revised{There is a large body of research that shows - as does our data - that privacy policies are not an appropriate medium for informing end-users about their privacy~\cite{BrunottePrivacyPlugin, PPLong2, PPUnderstanding2}. Rather, they are made \enquote{by lawyers for lawyers}. The fact that privacy policies are not read is not merely due to the users and their possible ignorance of their privacy~\cite{Karegar2020}. Our results indicate that end users might be interested~\cite{Cummings2021} and concerned~\cite{Anton2010} about their online privacy. Thus, alternative techniques such as privacy explanations can be an adequate solution to foster transparency on data practices.}

\subsubsection{Risk Beliefs}
\label{sec:rbresults}
The calculated risk scores for the risk beliefs metric ranged from 1.95 to 7.0 (M=4.78, SD=1.19). The histogram depicted in \figurename~\ref{fig:riskscores} shows an approximately normal distribution for the risk scores. The Shapiro-Wilk test confirmed that the risk scores are normally distributed (W=0.98, p=0.058). The majority of the respondents (68.4\%) have a risk score $> 4.3$. This suggests that respondents are not only aware of sharing their data online but also perceive it as a rather high risk.

We could observe that respondents with a higher risk score were more likely to read privacy policies with a positive correlation to Spearman's rank ($\rho=0.41,p<0.001$) and feel more uncomfortable using shopping portals ($\rho=0.52,p<0.001$), search tools ($\rho=0.51,p<0.001$), and games ($\rho=0.45,p<0.001$) (c.f. \figurename~\ref{fig:softwareusage}).

We also analyzed the possible relation between the IT experience level and the PSI. According to the PSI, the risk score among Privacy Fundamentalists is slightly higher on average (M=5.52, SD=0.91) than for Pragmatists (M=4.36, SD=1.07). The risk score for Privacy Unconcerned (three respondents) is the lowest on average (M=3.63, SD=1.85). This difference between groups is statistically significant with a negative correlation according to Spearman’s rank ($\rho=-0.49,p<0.001$).

However, according to our findings, Intermediates (M=5.01, SD=1.11) and Novices (M=5.13, SD=0.94) have a higher average risk score than Experts (M=4.55, SD=1.14) and Developers (M=4.77, SD=1.34). Arguably, these differences are lower than between the PSI groups. Possibly, the lower risk score among the Experts and Developers could be due to the fact that they have a deeper understanding of software systems (especially the developers). Thus, they \enquote{have an idea of what software does internally}. Intermediates and novices, on the other hand, see software systems primarily as black boxes (i.e., they know nothing about internal processes). This lack of knowledge could be the reason for the higher perceived risk. However, this assumption cannot be justified with our data since we did not ask any further questions in this direction.

\subsubsection{Privacy Threats and Concerns}
\label{sec:threatsandconcerns}
\begin{figure*}[t!]
	\centering
	\includegraphics[width=0.98\textwidth]{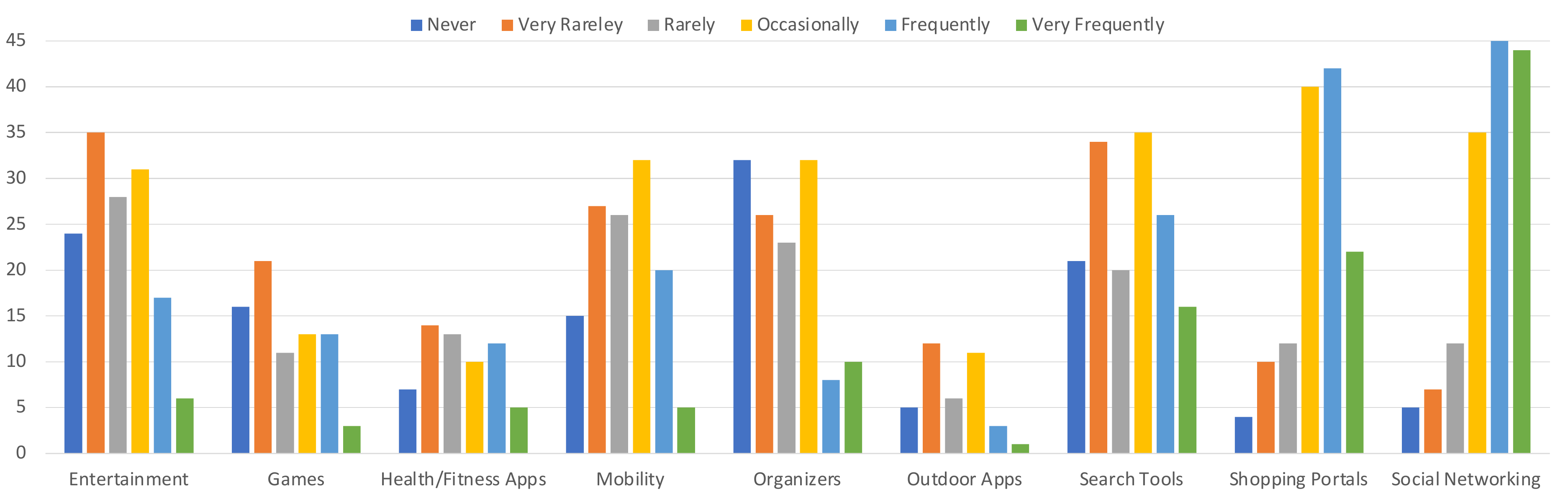}
	\caption{How often do you feel uncomfortable about privacy when using software or visiting websites related to these categories?}
	\label{fig:d1}
\end{figure*}

\para{Using Software.} In order to assess what kind of privacy concerns and threats are faced by the respondents, we asked how often they feel uncomfortable about their privacy, depending on the use of different software. For this purpose, respondents previously selected which of the software categories they use at all. In this way, we ensured that the respondents could only indicate how uncomfortable they feel with software that they actually use. The results are depicted in \figurename~\ref{fig:d1}.

\begin{figure}[]
	\centering
	\includegraphics[width=0.45\textwidth]{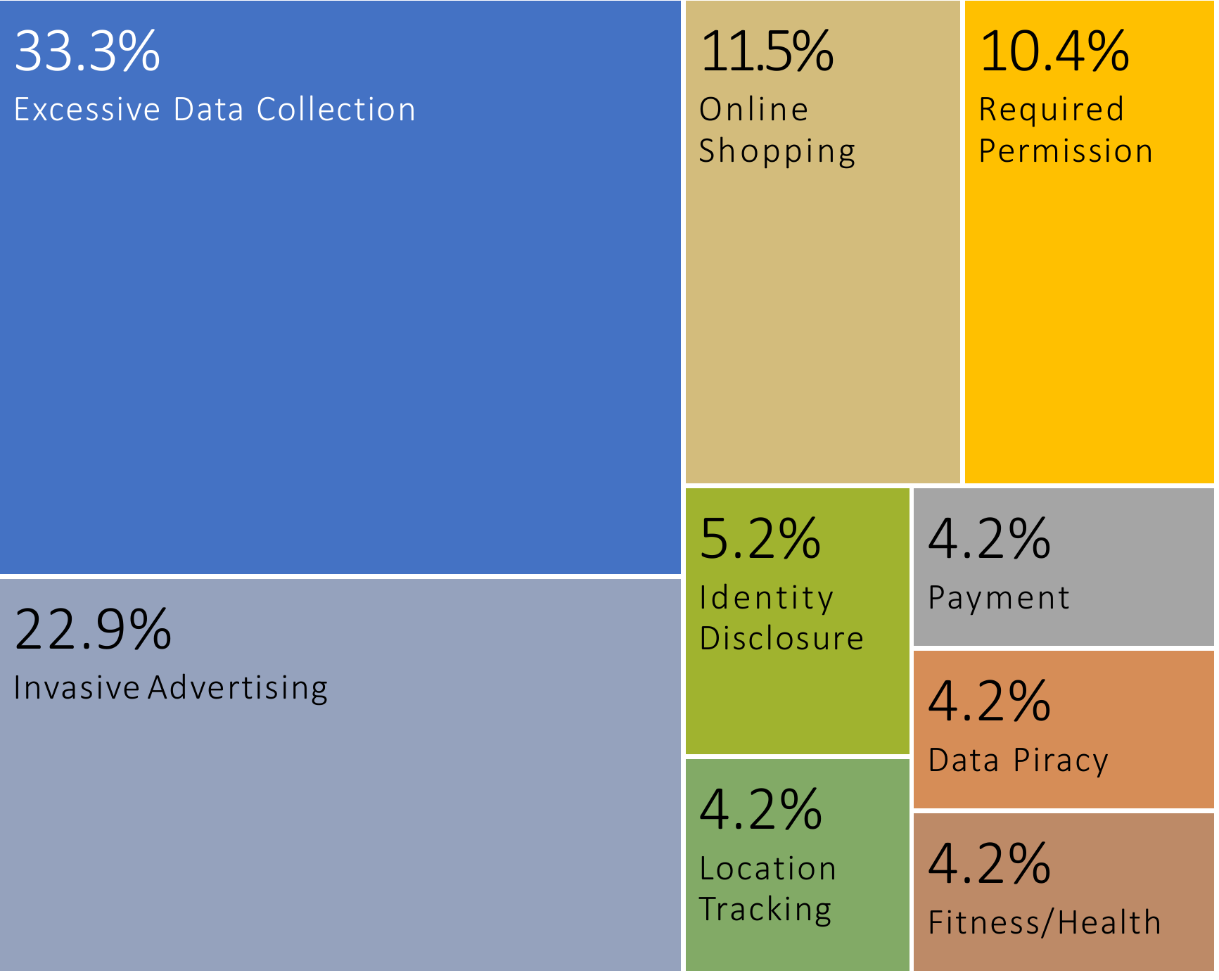}
	\caption{Categories where respondents felt uncomfortable while using software}
	\label{fig:d2}
\end{figure}

\para{Discomfort.} Respondents were asked to name a situation they had experienced in which they felt particularly uncomfortable using software in terms of privacy. By analyzing the respondents' answers, we were able to identify nine categories based on 96 codes. The categories are shown in \figurename~\ref{fig:d2}.

The greatest discomfort among respondents is triggered by \textbf{excessive data collection} (\emph{\enquote{it collects a lot of data about my everyday life}}), followed by \textbf{invasive advertising}. \revised{Invasive advertising is the practice of gathering and analyzing data and then presenting relevant ads based on that data. Respondents stated, for example, that they got the impression that offline conversations were being captured by technology such as smart-home assistants and that related advertising for products were being displayed during the next online search.} Others expressed discomfort about advertising based on the chat history of messaging services.

Many respondents related that they feel uncomfortable shopping online because the shopping portals often collect a lot of data, such as address data, payment data, order history, and consumption behavior. In addition, several respondents complained that many permissions are often required to execute software (camera access, location, etc.) without being clear what these accesses are needed for.

\begin{figure}[]
	\centering
	\includegraphics[width=0.47\textwidth]{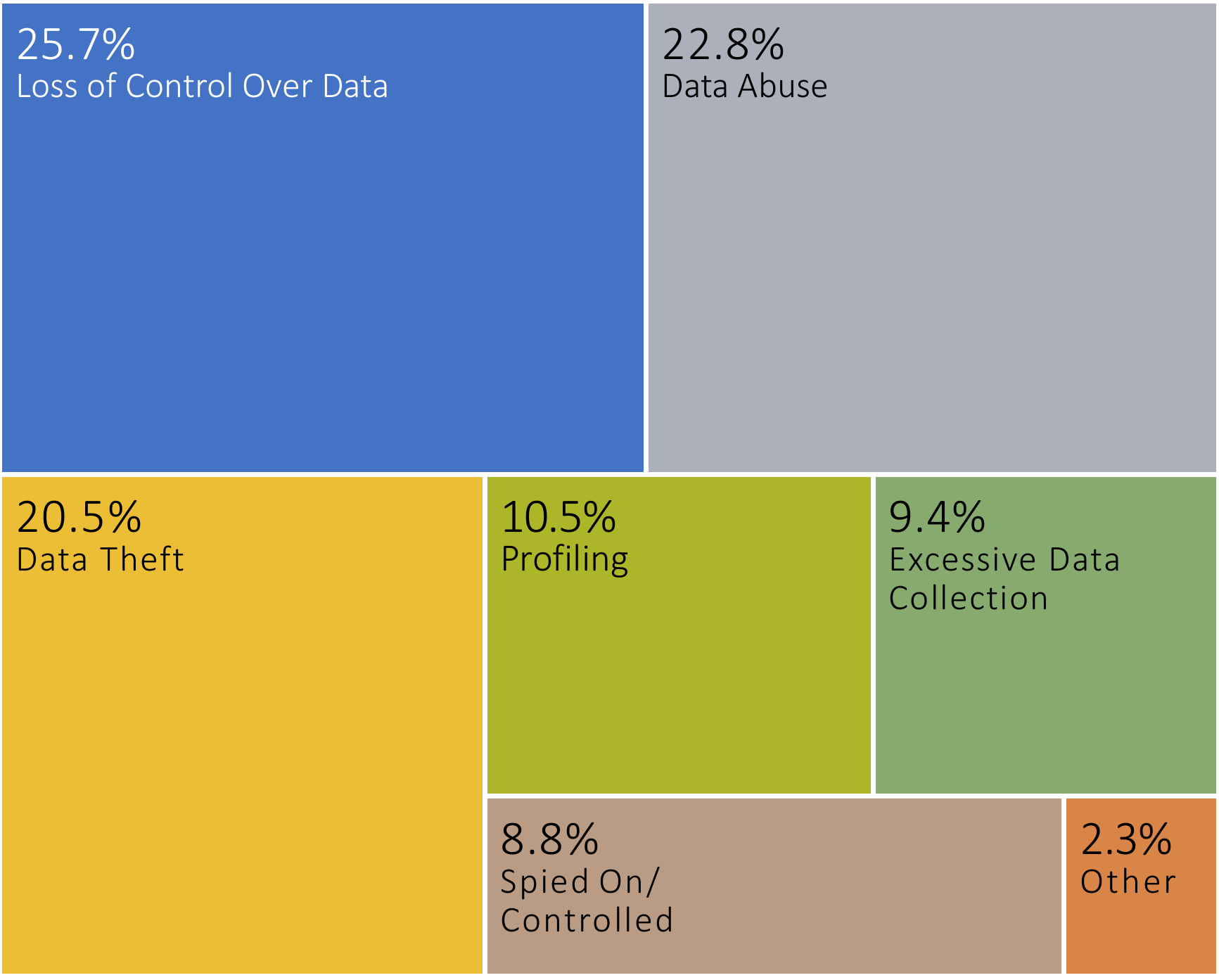}
	\caption{Privacy threats that respondents are concerned about}
	\label{fig:rbconcerns}
\end{figure}

\para{Privacy Threats.} We also asked respondents for one threat they are particularly concerned about regarding their privacy (see \figurename~\ref{fig:rbconcerns}). From the respondents' answers regarding their concerns, we were able to extract 171 codes. We then grouped these into seven categories. \textbf{Loss of control over data} is the concern most mentioned by respondents. Here, respondents expressed concern about not knowing who has access to their data, with whom this data is shared, and for what purpose this data is collected. Related concerns include \textbf{data theft} (\emph{\enquote{I am afraid that they will steal my bank details}}, \emph{\enquote{That one day our identities (like in the movies) can be stolen}}), \textbf{data abuse} (\emph{\enquote{Use of the information to commit crimes}}, \textbf{profiling} (\emph{\enquote{Creation and analysis of unique profiles via metadata consolidation}}). In addition, \textbf{excessive data collection} is also one of the concerns mentioned here, and some of the respondents are worried about being \textbf{spied on} (\emph{\enquote{Software spies on me}}). In the category \textbf{Other}, we categorized statements that consisted of only one word (e.g., \emph{\enquote{e-mail}}) or statements that would not fit into any of the other categories.

\begin{mybox}\textbf{Answering RQ1:} Taking into account the risk scores (Sec.~\ref{sec:rbresults}) and the concerns that our respondents expressed with respect to their privacy, we conclude that the majority of respondents have a high level of concern. Nevertheless, many respondents may be weighing the benefits and advantages of using certain services when it comes to their privacy. This could be supported by the PSI (60.65\% Privacy Pragmatists) as well as the fact that respondents often have concerns about using social networking tools and shopping portals but at the same time, use them frequently on a daily basis.
\end{mybox}

\subsection{RQ2 - Current Perception Regarding Privacy Explanations}
\label{sec:rq2}
\para{Interest in Explanations.} A hypothetical situation was presented to the respondents (Section~\ref{sec:research}). The goal was to analyze end users' need for privacy explanations in situations where a software system asks the user to disclose personal information. Based on this scenario, respondents were asked if they would be interested in a privacy explanation.

\begin{figure}[]
	\centering
	\includegraphics[width=0.45\textwidth]{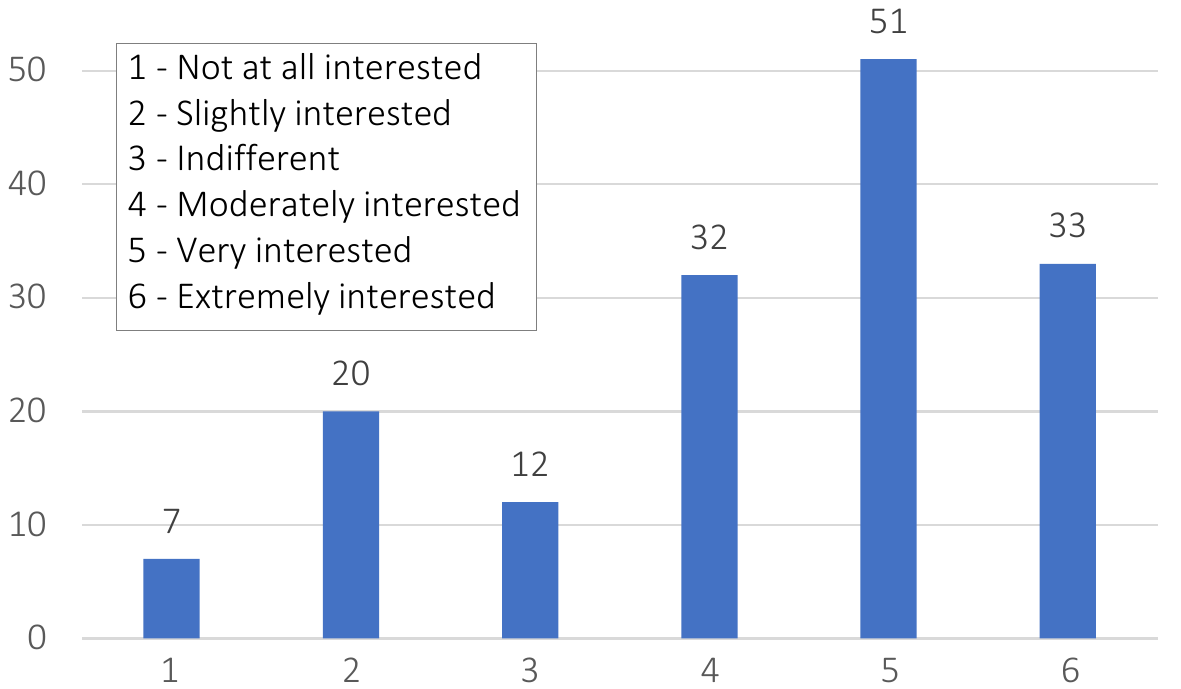}
	\caption{How interested are you in receiving a privacy explanation? (hypothetical scenario)}
	\label{fig:e1}
\end{figure}

87.7\% of the respondents are interested in receiving a privacy explanation (12.9\% slightly interested, 20.6\% moderately interested, 32.9\% very interested, and 21.3\% extremely interested), as shown in \figurename~\ref{fig:e1}. 

\begin{figure}[]
	\centering
	\includegraphics[width=0.45\textwidth]{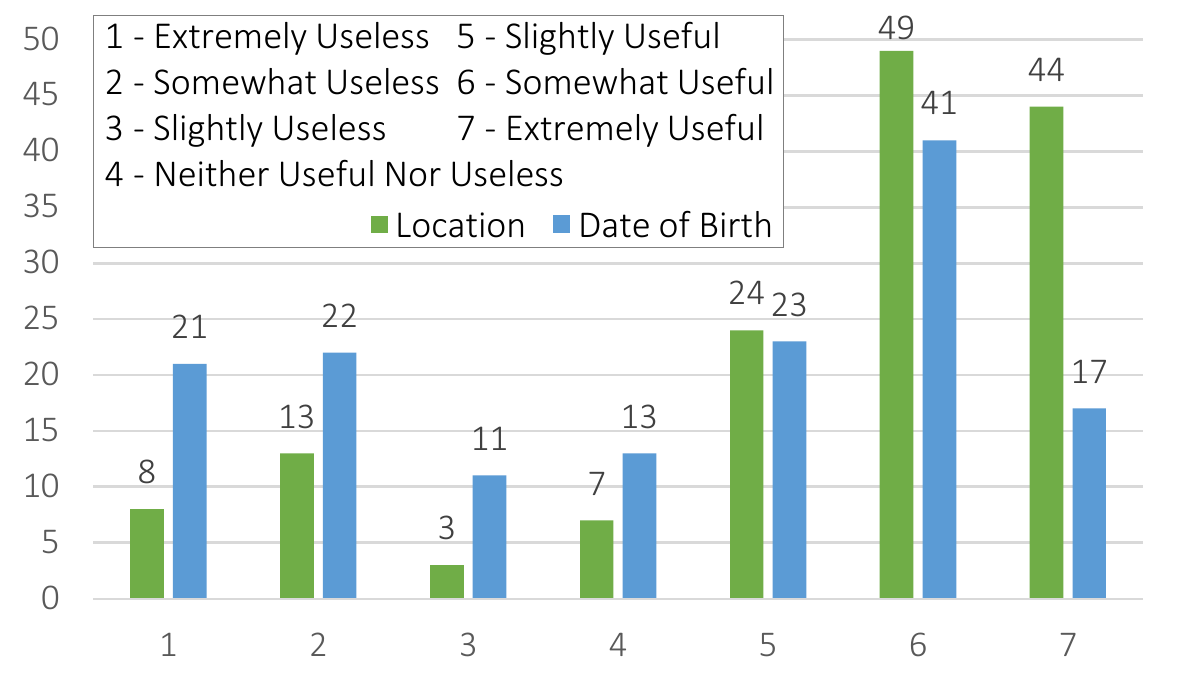}
	\caption{How useful do you find these types of privacy explanations? ($n=148$)}
	\label{fig:e1ae1b}
\end{figure}

For the respondents who wanted to receive an explanation regarding their privacy (95.5\%, $n=148$), we presented an explanation of how their data would be used with respect to the given scenario. According to Spearman's rank, we found that respondents with a higher risk score were statistically significant more likely interested in receiving a privacy explanation ($\rho=0.38,p<0.001$). The privacy explanation regarding the use of the location was: \emph{\enquote{in order to show you tours and recommendations near you, we need access to your location}} (E1). The explanation regarding the date of birth was: \emph{\enquote{based on your date of birth, we can show you recommendations of what other users your age have liked}} (E2). Respondents were then asked how useful they found each of these privacy explanations and whether the explanations helped them feel more comfortable about disclosing personal information.

\para{Usefulness.} \figurename~\ref{fig:e1ae1b} shows how useful respondents ($n=148$) found the privacy explanations. 79.1\% indicated that they found the privacy explanation regarding the location (E1) useful (\emph{slightly useful} to \emph{extremely useful}) and 16.2\% found them useless (slightly useless to extremely useless). The remaining 4.7\% were indifferent. Regarding the date of birth (E2) (\figurename~\ref{fig:e1ae1b}), 54.7\% perceived the explanation useful and 36.5\% found it was useless.

\begin{figure}[]
	\centering
	\includegraphics[width=0.44\textwidth]{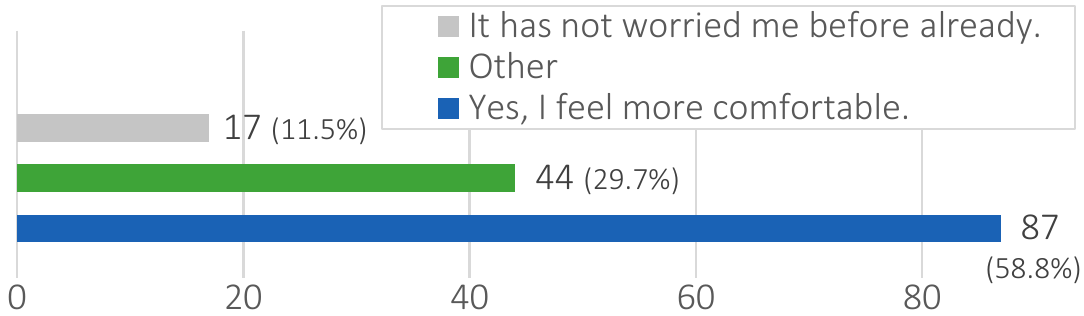}
	\caption{Do you feel more comfortable with the privacy explanations? ($n=148$)}
	\label{fig:e1b}
\end{figure}
\para{Well-being.} 
\revised{We asked if respondents felt more comfortable after receiving a privacy explanation. The results are shown in \figurename~\ref{fig:e1b}. Respondents who chose \enquote{other} as their answer had to justify their decision by entering an answer in the text box. The analysis of their answers resulted in 49 codes. We categorized the statements according to their meaning. The categories are \textbf{improvements} (14, 28.6\%), \textbf{criticism} (4, 8.2\%), \textbf{not sufficient} (14, 28.6\%), \textbf{mistrust} (9, 18.4\%), and \textbf{other} (8, 16.3\%).} 

The group \textbf{improvements} included statements such as \emph{\enquote{I would rather turn off the feature}} regarding the date of birth and \emph{\enquote{classification in age group would be sufficient}}. In the category of \textbf{cri\-ti\-cism}, statements such as \emph{\enquote{interests do not depend on age}} were included. We did assign statements such as \emph{\enquote{the reasoning on the age issue is not sufficient}} as well as \emph{\enquote{I would need to know in addition that these are the only reasons}} to the category \textbf{not sufficient}. Respondents' answers such as \emph{\enquote{not trustworthy}} as well as \emph{\enquote{feel spied on}} were assigned to the criticism category. Statements such as \emph{\enquote{the declaration is nonsense. I didn't want it.}} have been assigned to the \textbf{other} category. In the scenario presented to the respondents, the app asked for the date of birth. We deliberately constructed the scenario in such a way that instead of the year of birth - which would have been technically sufficient for an age recommendation - the date of birth was requested. Seven of the respondents explicitly mentioned this.

\begin{figure}[H]
	\centering
	\includegraphics[width=0.37\textwidth]{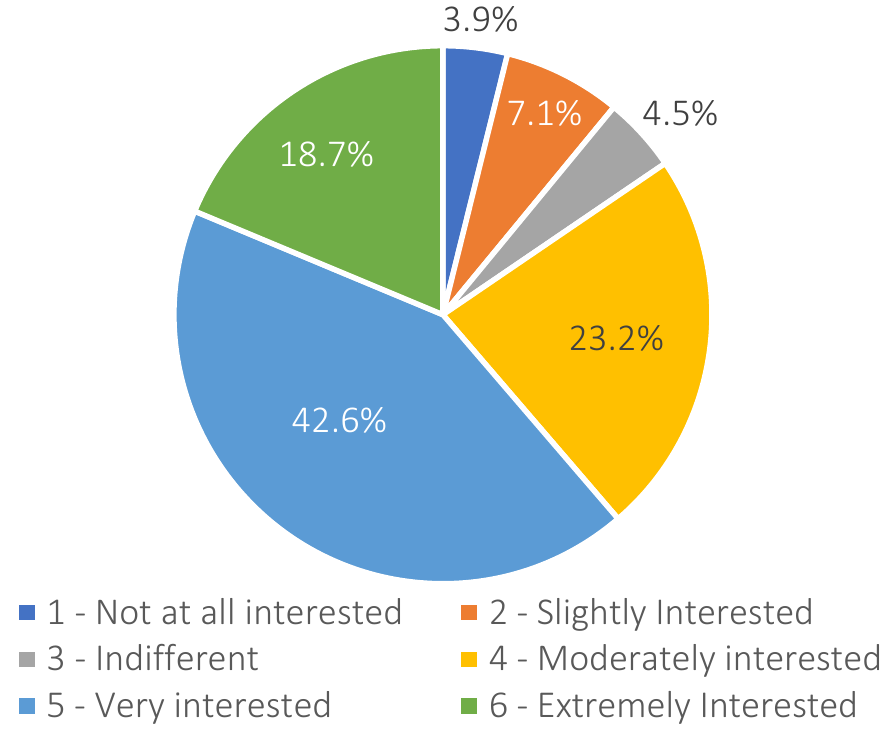}
	\caption{Are you generally interest in privacy explanations?}
	\label{fig:e4}
\end{figure}

\para{General Interest.} When asked whether respondents are generally interested in receiving explanations with respect to their privacy, the majority (91.6\%) of the respondents indicate that they are interested in privacy explanations (\emph{extremely interested} to \emph{slightly interested})\revised{, as shown in \figurename~\ref{fig:e4}}. When looking at respondents' risk score and their interest in receiving a privacy explanation, we get a similar picture as to the question from \figurename~\ref{fig:e1}: a positive correlation according to Spearman's rank ($\rho=0.29,p<0.001$). The higher the risk Score, the statistically significant higher the interest in a privacy explanation.

\begin{figure}[]
	\centering
	\includegraphics[width=0.45\textwidth]{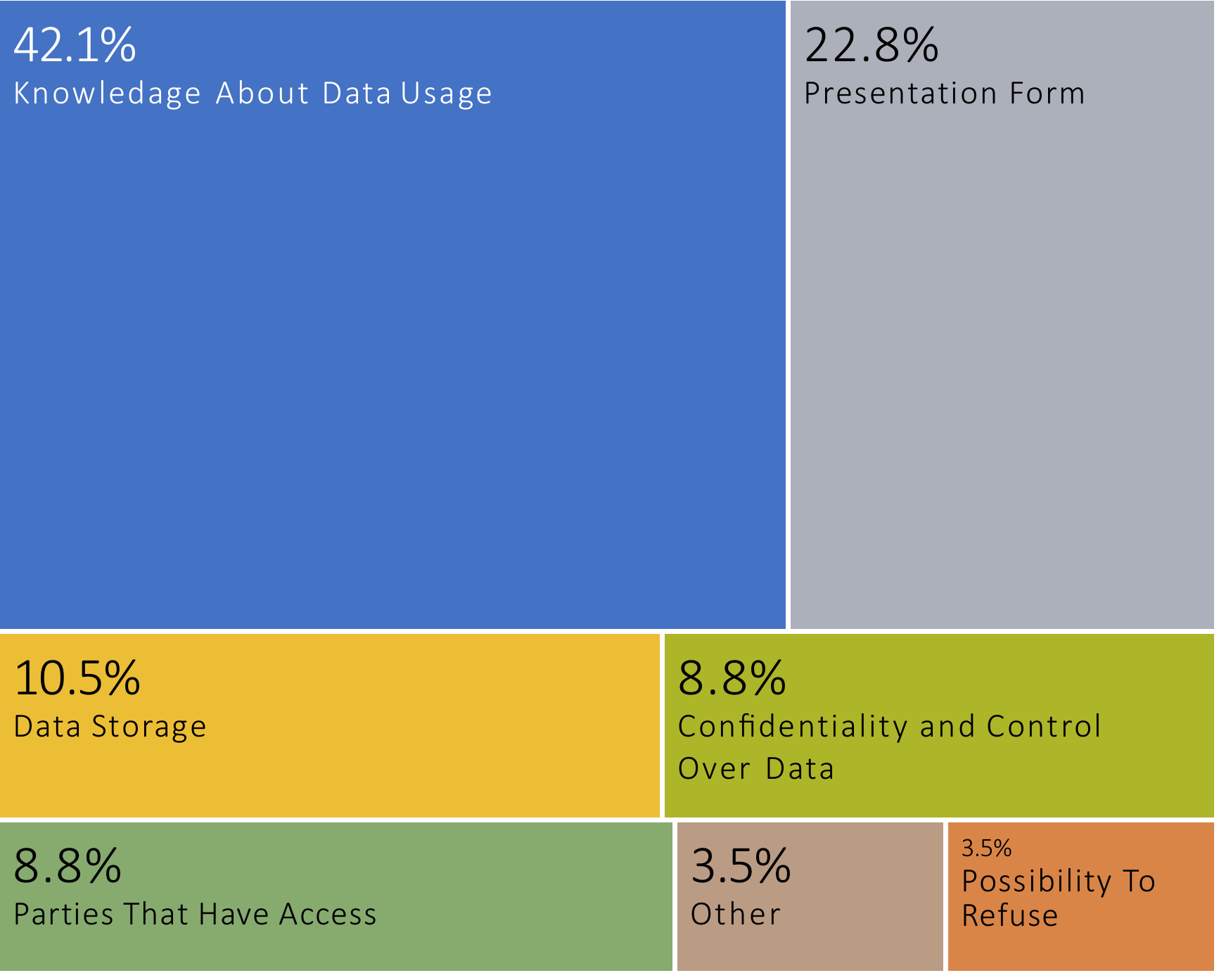}
	\caption{Important aspects to be considered in privacy explanations}
	\label{fig:e5}
\end{figure}

\para{Requirements on Privacy Explanations} In response to our open-ended question of what a privacy explanation should contain, respectively, what is  expected of it, the analysis of the data resulted in 57 codes and revealed the following picture, as shown in \figurename~\ref{fig:e5}.

First and foremost, end users want to understand how their data is being used. This includes knowledge about what data is used (42.1\%), why, and how. The \textbf{presentation form} also plays an important role (22.8\%). Respondents also indicated that explanations should be concise, precise, and written in simple language. Additionally, respondents said that icons can be visually supportive.

Respondents also expressed that they would like to know where the data is stored, for how long, how it is protected, and if or when the data is deleted. We clustered statements of this type in the \textbf{data storage} (10.5\%) group. Respondents want to \enquote{\emph{be sure that this data cannot be sold to other companies}}. They want to be assured that the data will be kept confidential and \enquote{\emph{not used for anything else}} (\textbf{confidentiality and control over data}, 8.8\%). Respondents also indicated that privacy explanations should provide information about \textbf{parties that have access} (8.8\%) to the data and  give information about a \textbf{possibility to refuse} (3.5\%). The category \textbf{other} comprises statements for which we were not able to make a relation to our question.

\begin{mybox}\textbf{Answering RQ2:} The vast majority of respondents (91.6\%) are interested in privacy explanations and consider them useful since they inform them about data practices. For this purpose, it is important for them to know what the data is used for and how it is stored. The presentation form of such an explanation also plays an important role. A privacy explanation should be easy to understand and be connected to the user's present context. That means, for example, if an app requires a user's location, the app should explain why the location is needed.
\end{mybox}

\subsection{RQ3 - Privacy Explanations and the Concept of Trust}
\label{sec:rq4}
To answer RQ3, we asked respondents to name up to three benefits they think may be associated with privacy explanations. Following this, we asked when they should be presented and whether respondents agree that privacy explanations can help increase the level of trust in a software system.

\para{Benefits.} 137 respondents answered this open-ended question, which resulted in 135 valid responses. Each valid response was analyzed and resulted in a total of 363 codes.

\begin{figure*}[htbp]
	\centering
	\includegraphics[width=0.98\textwidth]{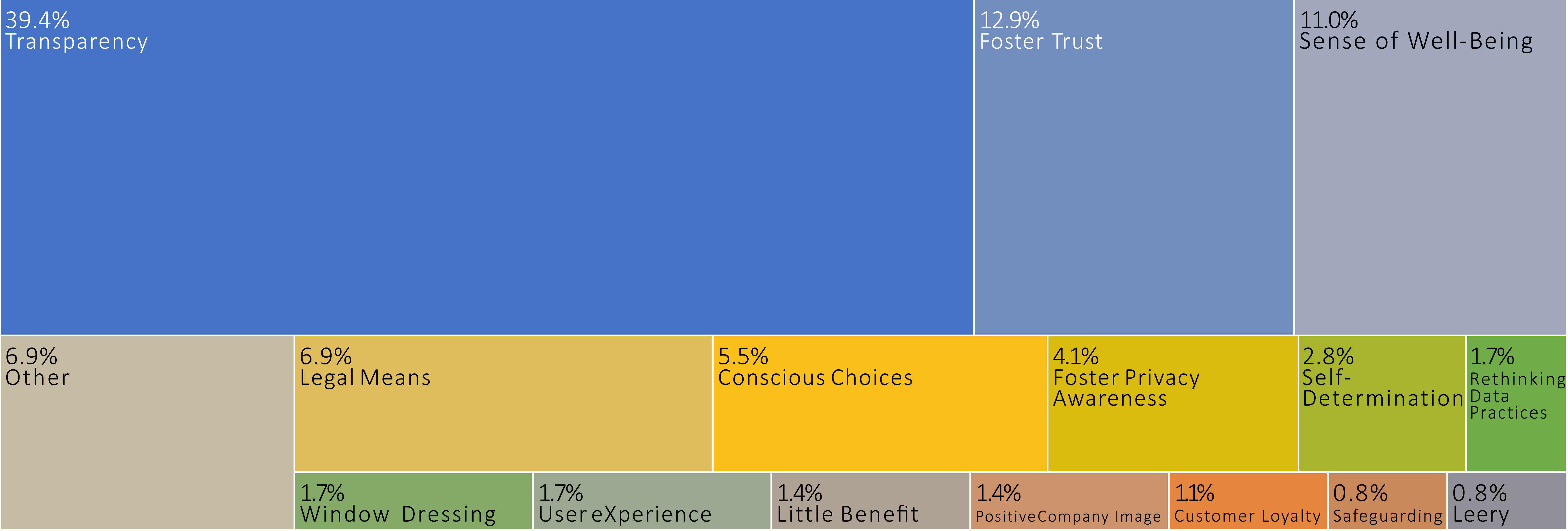}
	\caption{What do you think is the benefit of privacy explanations?}
	\label{fig:f1}
\end{figure*}

We summarized 24 codes (6.6\%) in the category \textbf{other}, which we could not assign to any advantage. These statements could not have been assigned to a disadvantage either, because they partly showed a lack of understanding on the part of the respondents (\emph{\enquote{data sale}}, \emph{\enquote{minimizes spam}}, \emph{\enquote{the company makes money with it}}). Since we would have to interpret too much into such statements and a classification into another category would be too subjective, we decided to classify such statements into this cluster.

\textbf{Transparency} is the largest category and 39.4\% (143) of the 363 codes were grouped here. Statements such as \emph{\enquote{transparency with the user}}, \emph{\enquote{clarity}} or \emph{\enquote{increases users understanding of the software they are using}} were included in this category.

The second largest cluster is \textbf{foster trust} (47 codes, 12.9\%). Respondents mentioned benefits such as \emph{\enquote{strengthen trust}}, \emph{\enquote{build trust towards the system}}, and \emph{\enquote{more trust in the application and the company}}. In addition to trust-fostering benefits of privacy explanations, respondents expressed that explanations can also create a \textbf{sense of well-being} (40, 11.0\%). In this category, we have grouped statements such as \emph{\enquote{reduces uncertainty}}, \emph{\enquote{transmits a sense of confidence}}, and \emph{\enquote{improves users' sense of well-being}}.

In addition, according to the respondents, privacy explanations contribute to make \textbf{conscious choices} (17, 4.7\%) when disclosing personal data. They state that an explanation \emph{\enquote{allows users to give informed consent}}, or \emph{\enquote{one can decide more consciously whether it is worthwhile to disclose the data}}. According to the respondents, \textbf{self-determination} (10, 2.8\%) might be a further benefit of privacy explanations because it gives \emph{\enquote{control over your own data}}. We have grouped statements such as \emph{\enquote{helps people to think twice}} as well as \emph{\enquote{becoming more aware of what happens to data}} in the category \textbf{foster privacy awareness} (15, 4.1\%).

In addition to these categories, respondents also see a benefit from a legal perspective. Statements such as \emph{\enquote{comply with Federal legislation}} or \emph{\enquote{compliance}} were grouped into the category \textbf{legal means} (25, 6.9\%). Other benefits of privacy explanations considered by respondents include a \textbf{positive company image} (5, 1.4\%), \textbf{customer loyalty} (4, 1.1\%), and a \textbf{rethinking of current data practices} (6, 1.7\%). Here they expressed privacy explanations \emph{\enquote{force a company to think about what data to collect}} or \emph{\enquote{the operators inflict a usage policy upon themselves, which sets a boundary}}.

In addition, respondents indicate that privacy explanations foster \textbf{user experience} (UX) (6, 1.7\%) as well as that they may safeguard a person's interests (\emph{\enquote{protection of own interests}}) or the person themselves (\emph{\enquote{protection of the own person from, e.g., unnecessary advertising}}). We coded statements like the last two as \textbf{safeguarding} (3, 0.8\%).

While some of the respondents see only \textbf{little benefits} (5, 1.4\%), others also expressed criticism. They described privacy explanations as \textbf{window dressing} since they could (\emph{\enquote{fool the users}} and may cause \emph{\enquote{Consumer confusion (appeasement/downplay)}}). Other statements suggest that respondents are rather leery of privacy explanations. We have therefore assigned statements such as \emph{\enquote{you can not trust that this will be adhered to}} to the \textbf{leery} (3, 0.8\%) category.

\begin{figure}[H]
	\centering
	\includegraphics[width=0.3\textwidth]{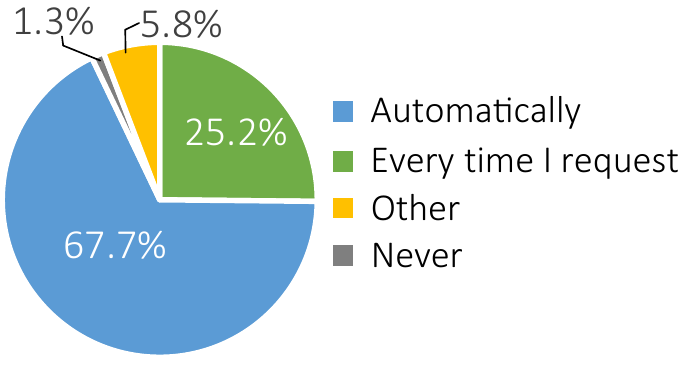}
	\caption{When should a privacy explanation be presented?}
	\label{fig:f2}
\end{figure}

\para{When to Present.} We asked respondents when privacy explanations should be presented. 
\revised{The results are shown in \figurename~\ref{fig:f2}}.

The analysis of these answers resulted in 18 codes. We have divided the statements into the following categories according to the answer options \emph{automatically} (6, 31.5\%), \emph{on change} (6, 31.5\%), and \emph{on request} (6, 31.5\%). \revised{5.8\% of respondents chose \emph{other} as an answer option and provided their answer in text form.} The analysis of the answers reveal that respondents want an explanation presented automatically, and every time something changes, but also whenever they request an explanation.

\begin{figure}[]
	\centering
	\includegraphics[width=0.45\textwidth]{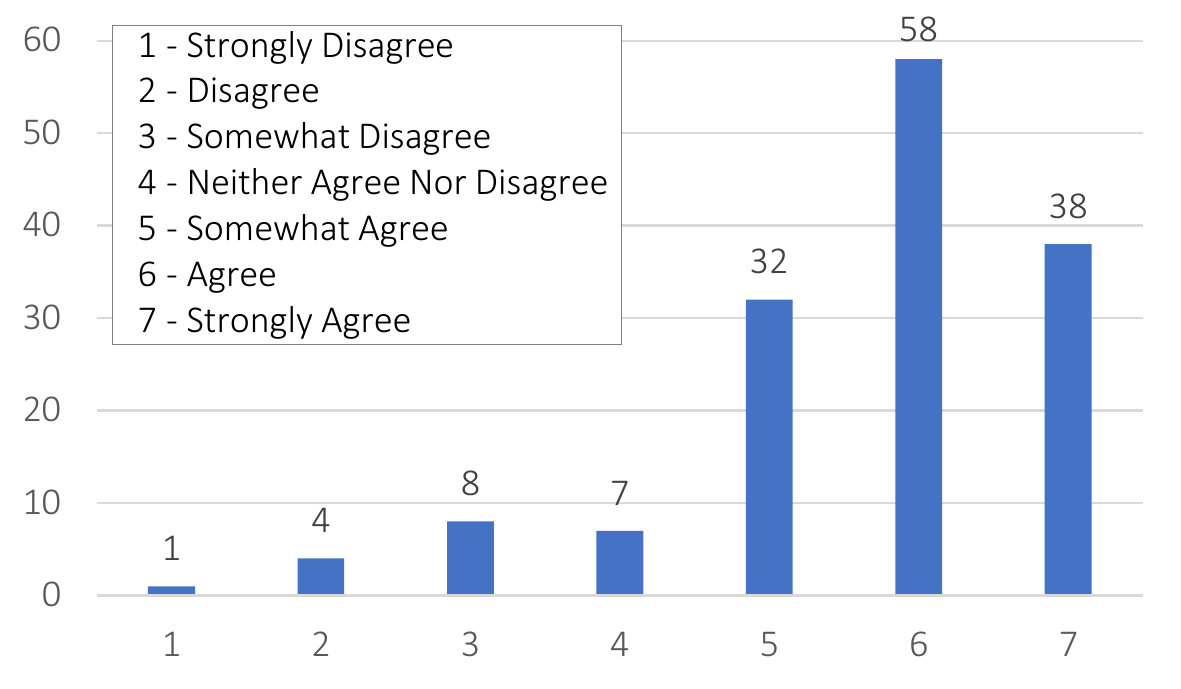}
	\caption{Can privacy explanations be a possible factor to increase the level of trust in a software system? $(n=148)$}
	\label{fig:e1c}
\end{figure}

\para{Privacy Explanations and Trust.} 
We asked our respondents $(n=148)$ whether they agree that privacy explanations can be a possible factor to increase the level of trust in a software system. We received a clear picture here, as shown in \figurename~\ref{fig:e1c}, which is in line with responses from \figurename~\ref{fig:f1}. 

\begin{mybox}\textbf{Answering RQ3:} Our results suggest that privacy explanations might foster trust as well as privacy awareness. They can provide more transparency with respect to the applied data practices of a system and enable end users to make more conscious choices regarding their privacy  (\figurename~\ref{fig:f1}).

Privacy explanations can make end users feel more secure and increase their well-being while operating a software system. Furthermore, privacy explanations might have a positive impact on the trustworthiness of a software system and, in turn, foster end user trust in the system.
\end{mybox}

\section{Discussion}
\label{sec:discussion}
\revised{Our results led to six findings that helped to shape four concepts. In this section, we discuss these concepts and the related findings. The first concept explores the end users' perception on privacy explanations and the privacy paradox, discussed in \subsectionautorefname~\ref{sec:diss-perception}. The second concept explores the relationship between trust and privacy explanations and introduces the trust-coin metaphor, discussed in \subsectionautorefname~\ref{sec:diss-trust}. The third concept explores the requirements on privacy explanations, discussed in \subsectionautorefname~\ref{sec:diss-reqs}. And the fourth concept explores the relationship between relevant stakeholders for privacy-aware systems, discussed in \subsectionautorefname~\ref{sec:diss-trialog}.}

\subsection{\revised{Users' Perception on Privacy Explanations}}
\label{sec:diss-perception}

Our findings show that participants' perceptions of privacy explanations are paradoxical: many participants do not actively take action about data practices, despite the fact that they see privacy and privacy explanations as desirable and beneficial. Two findings enable us to more fully grasp the subtleties of the potential end users' perception of privacy explanations:

\para{Finding 1:  Privacy explanations are beneficial.} According to \revised{our results, a vast majority (91.6\%)} of the respondents are generally interested in privacy explanations. A closer look at the explanations given \revised{in the hypothetical scenario} and the respondents' reactions to those explanations reveal that they perceive privacy explanations as supportive and feel more comfortable (\figurename~\ref{fig:e1b}) when they receive information about data usage.

\revised{74.3\% (58.8\% with respect to E2)} of the respondents who received an explanation felt more comfortable after knowing about how personal data would be used. Respondents indicate that explanations reduce their uncertainty, provide more security and confidence, which all, in turn, result in a feeling of well-being.

Informing the user \emph{(a)} that personal data about them is being collected, \emph{(b)} what data is being collected, and \emph{(c)} how this data is being used, contributes to privacy awareness, as described in~\cite{Poetsch2009}. At the same time, being informed enables self-determined and conscious decisions when using software systems, which was also mentioned by the respondents. Following this line of thought, it also becomes clear why some of the respondents consider privacy explanations as a kind of \emph{safeguard}. A safeguard between one's own \emph{online privacy} and the uninformed/unintentional disclosure of privacy aspects.

\para{Finding 2: The privacy paradox.} To evaluate the privacy attitudes of our respondents, we grouped them according to the PSI items (Section \ref{sec:psi}), calculated their risk beliefs (Section \ref{sec:rbresults}), and asked them what concerns they have in terms of privacy (Section \ref{sec:threatsandconcerns}). 

\revised{Even though respondents are concerned and care about their privacy they avoid engaging in the necessary privacy behavior.} This phenomenon is known in the literature as \textbf{privacy paradox}. This term was coined by Barnes~\cite{Barnes2006} and is well researched by many others~\cite{GERBER2018, KOKOLAKIS2017, Hargittai2016, Bandara2020, Poetsch2009}. In a nutshell, the privacy paradox states, \enquote{I am aware that my privacy is being violated, yet I continue to utilize this service}. The privacy paradox affects all generations~\cite{Pentina2016} and \enquote{cannot be attributed solely to either a lack of understanding of or a lack of interest in privacy}~\cite{Hargittai2016}, as reflected in our results.

Furthermore, according to Hargittai and Marwick~\cite{Hargittai2016}, end users' lack of appropriate privacy behavior is rooted in the apathy they developed towards online privacy since the systems are often black boxes, have opaque data practices, and the privacy controls available change frequently. As a result, it is difficult and confusing for end users to comprehend how their personal data flows. This leads to frustration and worries, which yields self-censorship and apathy. \revised{Therefore, easy accessible information should be available to inform end users about data practices.}

This paradoxical attitude of end users~\cite{GERBER2018, Mourey2020} is one aspect that conveys the complexity and challenges of dealing with privacy and is also reflected in our results. This complexity stems from the fact that users weigh costs and benefits when making decisions~\cite{Earp2003}. \revised{Different user attitudes} also reflect on privacy behavior~\cite{PrivacyAUsefulConcept, Rudolph2018}. For instance, some users are more concerned or \enquote{leerier} than others. This phenomenon was also evident when we asked about the benefits of privacy explanations, where some of the respondents expressed their concerns and worries. \revised{These concerns also reflect an important point related to privacy: \emph{trustworthiness}. The next \subsectionautorefname~\ref{sec:diss-trust} will explore trust and trustworthiness in more detail.}


\subsection{\revised{The Trust-Coin - Trust and Trustworthiness}}
\label{sec:diss-trust}

\revised{Trust in IT is an important concept since today's society heavily relies on IT. Trust in the competence of a system means that a system has the functionality or functional capability to perform a particular task that the trustor requests~\cite{McKnight2005}. In terms of privacy, opaque, incomprehensible data practices, and lack of transparency can harm this trust. Our participants' responses also confirmed this.}

\revised{To better understand trust and trustworthiness, let us imagine trust as a coin: a \enquote{trust-coin}. On the one side of the coin is the end-user trust, which represents their perception of trust towards a system. On the other side of the coin is trustworthiness as a property or quality aspect of a system. In terms of privacy, we cannot look at the two sides separately because they are interwoven. This means that if software engineers want end users to trust their system, they should ensure that their system \emph{is} trustworthy in terms of privacy.}

\para{Finding 3: Privacy explanations as a means to trust.} According to our results, privacy explanations are a means to increase the level of trust in software systems (\figurename~\ref{fig:e1c}). Our results suggest that they might help to establish a relationship of trust between the end  user and the system by increasing data transparency and clarity. Privacy explanations might help to put the user in control so that they can make self-determined, conscious choices with respect to their personal data. Respondents' answers reveal what requirements on privacy explanations (\figurename~\ref{fig:e5}) can serve respondents' privacy concerns. 


\subsection{\revised{Requirements for Privacy Explanations}}
\label{sec:diss-reqs}
To incorporate privacy explanations into systems, stakeholder requirements need to be elicited. It is important to meet the needs and expectations that stakeholders have regarding such explanations, as evidenced in \emph{Finding 1}. Otherwise, they may defeat their purpose of informing the end user regarding their privacy or even cause mistrust~\cite{papenmeier2019model, pieters2011explanation}.

\revised{In light of this, it is not enough to provide \enquote{any} explanation. A privacy explanation must make sense to the end user, considering their individual characteristics and context~\revised{(c.f. explanation E1 and E2 in \figurename~\ref{fig:e1ae1b}).}} 

\revised{This provides important insights for the development of explainable and privacy-aware systems. End users demand data reduction and data economy because their privacy is important to them, as mentioned above. That means for the design of the systems, data economy, responsible, and fair use of personal data is required~\cite{Koskinen2019, Schafer2017}. For our scenario, this means that instead of asking for the date of birth, this system should only ask for the year of birth. Taking into account the respondents' suggestions for improvement, an explanation should be able to provide further information on demand. For example, what the consequences are for the user if they provide their consent.}

\begin{figure}[H]
	\centering
	\includegraphics[width=0.47\textwidth]{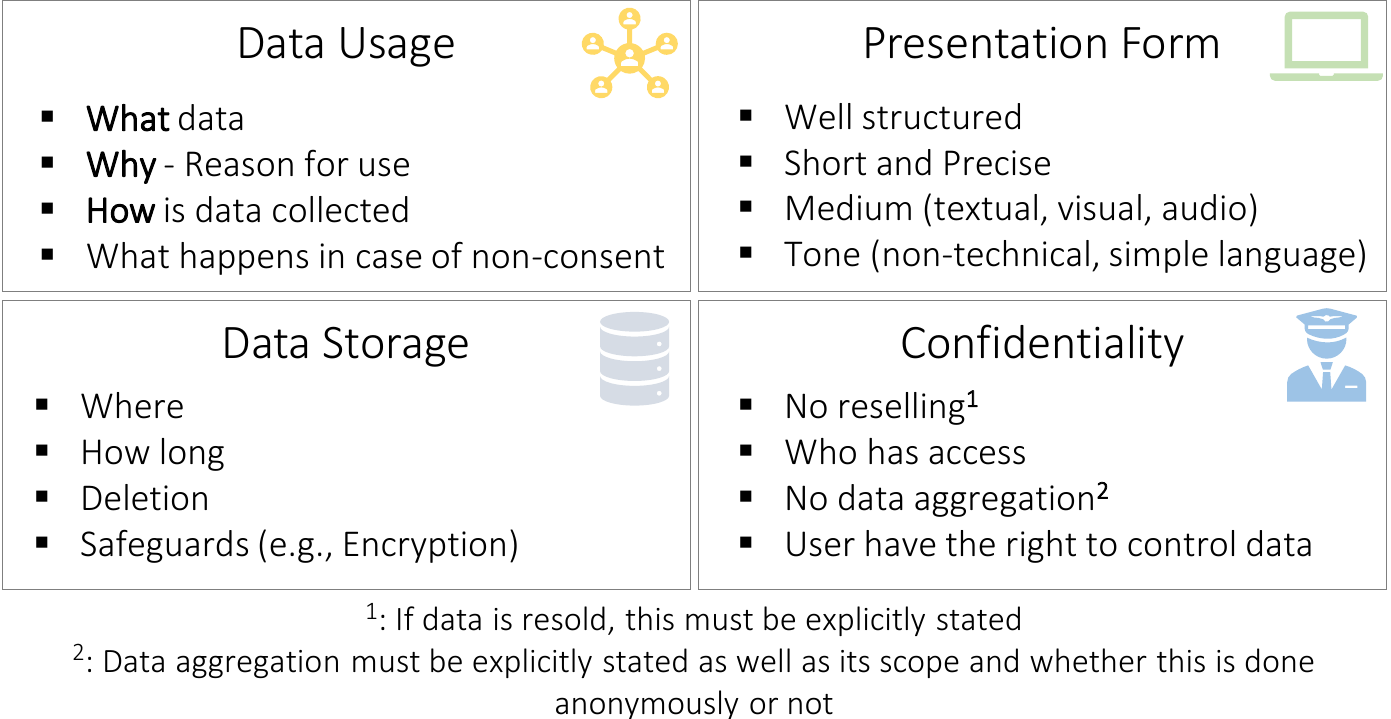}
	\caption{Requirements for privacy explanations}
	\label{fig:pereqs}
\end{figure}

\para{Finding 4: The \enquote{trust} side of the coin.} We asked respondents what they expect from a privacy explanation or what it should contain (\figurename~\ref{fig:e5}). This helped us to identify aspects that should be considered in privacy explanations, according to the answers in our survey \revised{and in line with our definition of online privacy (Section~\ref{sec:onlineprivacy})}. These aspects can work as high-level requirements for privacy explanations that should be met, as depicted in \figurename~\ref{fig:pereqs}. We categorized these high-level requirements into four categories. These first requirements \revised{in conjunction with our definition of online privacy} can serve as a starting point for software engineers to understand what elements should be considered when designing privacy explanations. We list each one of the four categories below.

\textbf{Data Usage:} Information about the use of personal data is crucial to clarify \textbf{what} data is used, \textbf{why}, and \textbf{how} it is used. We refer to this as the \textbf{2W1H} principle. A privacy explanation should inform the user about it.

\textbf{Data Storage:} Users should know where and how long the data is stored. An explanation should provide information whether the data is temporarily collected or in the long-run, as well as what \textbf{safeguards} are put in place to prevent accidental or deliberate privacy violations.

\textbf{Confidentiality:} Confidentiality is also an essential point in terms of privacy. This includes disclosing who has access to the data and whether the data is resold or used for further aggregation. This includes opt-out mechanisms (e.g., for reselling data) to ensure that users retain control over their own data.

\textbf{Presentation Form:} Privacy explanations should be well-structured, short, and precise, as well as communicated in a non-technical and simple language. 

\revised{These requirements also reflect what we claim in our proposed definition of online privacy (\sectionautorefname~\ref{sec:onlineprivacy}). What privacy aspects an individual is willing to share with others corresponds to the data usage and confidentially categories. The technical implementation of these categories is realized via the requirements for presentation form and data storage.}

Privacy explanations should be seen as a \textbf{context-aware, usable privacy feature}.  According to our findings, a system should provide such explanations in the relevant  context (automatically and/or on request when certain personal data are used) and be able to react and (re-)enter into a dialog with the end user, in case a policy changes.

\para{Finding 5: The \enquote{trustworthy} side of the coin.} In order to gain the trust of end users, systems must be reliable, i.e., trustworthy in terms of privacy. Therefore, systems that encompass quality aspects such as accountability, fairness, and ethics~\cite{Koskinen2019, Rantanen2019} are needed as well as where privacy is the \enquote{default setting}~\cite{Cavoukian2009}. In addition, these systems must implement applicable laws and regulations in terms of privacy in order \enquote{to protect individual privacy in information processing}~\cite{Bowman2015}.

With regard to privacy explanations, this means if a system explains to a user for what purpose a certain privacy aspect is needed, the system must not use it for any other purposes. The system must guarantee this in order to count as trustworthy. \revised{By giving explanations, the system can support the end user in building trust in the system itself. This also illustrates why it is important not only to distinguish between trust and trustworthiness, but also to specifically take both quality aspects into account.}


\subsection{\revised{Trialog of privacy -- Keep it User-Centric}}
\label{sec:diss-trialog}
\enquote{Privacy requires a dialogue between two types of people: those who speak policy and those who speak engineering}~\cite{Ohm2015}. This means that policy-makers (e.g., legislators and lawyers) must work together in direct dialog with software engineers and support each other in eliciting privacy requirements, assuring that they are legally compliant, and translating them into systems, regulations, and norms. According to Ohm, lawyers respond and react to what engineers engineer instead of communicating with each other~\cite{Ohm2015}.

\begin{figure}[]
	\centering
	\includegraphics[width=0.4\textwidth]{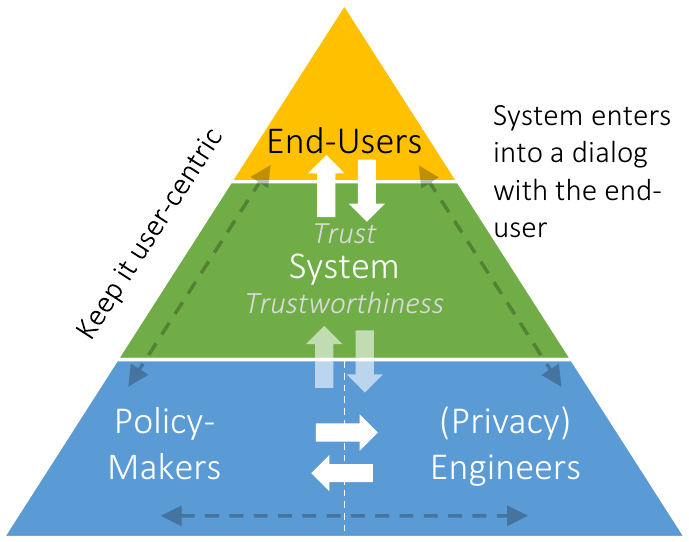}
	\caption{Trialog of privacy}
	\label{fig:trilog}
\end{figure}

From our point of view, the user, who should actually be the focus (be the center), takes a back seat in Ohm's statement. We propose a \textbf{trialog of privacy} (see \figurename~\ref{fig:trilog}). Therefore, we enhance Ohm's statement and suggest: 

\begin{mybox}
Privacy requires a trialog between three types of people: those who are end users \revised{and whose privacy is at stake}, those who speak (privacy) engineering, and those who speak policy.
\end{mybox}

\para{Finding 6: Relevant stakeholders for privacy-aware systems.} According to \figurename~\ref{fig:trilog}, the foundation for a privacy-aware system lies in the dialog between policy-makers and privacy engineers. We adopt the term privacy engineer in reference to Bowman~\cite{Bowman2015} who stated that  \emph{privacy is not something that can be fully addressed with a few architectural decisions made in the design phase alone}. The commitment to privacy is dynamic. As technology grows and is adopted by more users in different contexts, this commitment needs to be maintained. Therefore, Bowman postulates the role of a privacy engineer who maintains and is responsible for the privacy architecture~\cite{Bowman2015}. Similar to the role of a usability engineer who is responsible for the usability architecture. 

When these two parties (policy-makers and privacy engineers) are in dialog, the system might be built on a privacy-aware and trustworthy basis. To consider the end user as a third party and complement the trialog the system should \enquote{consciously designed around the interests and needs of individual users}~\cite{Cavoukian2009} in order to meet  their individual privacy preferences and enter into a dialog with the users via  explanations. This is what is meant by \textbf{keep it user-centric}\footnote{\emph{Respect for User Privacy -- Keep it User-Centric:} 7th Foundational Principle of Privacy by Design by Cavoukian~\cite{Cavoukian2009}} and why these three parties should be counted among the group of relevant stakeholders.

\section{Threats to Validity}
\label{sec:threats}
The strategy to select the participants has some limitations. Despite the fact that we received answers from different countries of the world, the majority of responses came from Germany and Brazil. This may not reflect the whole population and may threaten the global generalizability of our results. Although 155 participants provided a substantial amount of responses, some of the conclusions might be affected by this size and should not be overgeneralized. Most of the respondents of our study have  profound IT knowledge. Our population may not take into account people who have difficulties operating software systems. Therefore, we can not generalize the needs regarding privacy explanations, but we get an overview of what different people think.  
For RQ2 we identify different concerns while using applications. We only can evaluate the answers of our respondents and it might be that there are much more reasons of concerns for privacy, e.g. people who don't have IT knowledge. The other findings for our research have the same limitations. To find more concerns further experiments need to be conducted. To mitigate the thread that the analysis is too subjective we use \textit{in vivo coding} by two researchers. Each of them categorizes the data. During the second cycle they discuss and compare their findings to increase the consistency and reliability.   

To evaluate the need for privacy explanation we use a hypothetical scenario. This scenario is potentially not the daily use for the participants, but it might encounter them in real life. Only users of smartphones could better empathize with this situation. This situation confronted the respondents only in a scenario of vacation. The results could be different when they will confront in a business or financial scenario, because it could have posed a greater threat to their privacy.

Another aspect is that a good question wording and instrumentation layout are crucial for the results of a survey. We followed guidelines and conducted pilot-tests to ensure these aspects. However, the order of questions in the questionnaire may have impacted in the participants’ understanding about whether we were asking questions about the need to receive explanations in a general context or related to the previous question about a more specific context. However, we considered that this would be helpful for participants who could have difficulties to imagine other situations where they would need explanations. 

We decided to disclose our online questionnaire and raw data\footnote{\textbf{Note for the reviewers:} in case of acceptance of this article, we will publish the data on Zenodo for free access.} so that other researchers may be able to replicate and comprehend how we have drawn our conclusions and recommendations from the data. This step should serve as a final strategy to mitigate threats to the internal validity~\cite{SurveyMaterial}.

\section{Future Directions}
\label{sec:futurework}
Our study provided us with valuable insights into how respondents perceive privacy explanations and that these can make an important contribution to communicating data practices to end users in an comprehensible and transparent way.

Building upon findings of our research, we need to investigate how to translate our set of requirements for privacy explanations into a system. To keep privacy explanations \emph{as simple as possible} in order not to overwhelm the user, we plan a user study in which we survey how privacy explanation must be engineered~\cite{Brunotte2022}. We assume, that an hierarchical information structure may be beneficial. In the first place, a system informs a user about personal information usage, as happened in our scenario. In a next step, the system shall provide further information to the end users upon request, according to their individual privacy preferences and our 2W1H principle. Overall, it is important that it follows an actionable and operationalizable process.

Finally, we suggest more collaboration across disciplines (humanities, law, and  computer science) since \enquote{privacy is not an individual process, but rather a collective effort that requires cooperation} (\cite{Hargittai2016}) of those who are involved. Research could address how the trialog of privacy could be integrated into existing privacy-frameworks~\cite{Awanthika2017, Notario2015} that are currently focused on the engineering part.

\section{Conclusion}
\label{sec:conclusion}
In this article, we conducted an online survey with 155 participants to investigate end users perception and attitude towards privacy explanations. 

Our findings suggest, that end users perceive privacy explanations as beneficial and they  may influence the well-being of end users. 91.6\% of our respondents are generally interested in receiving such explanations. We found  that privacy explanations may also be seen as a means toward end user trust \revised{ and aim to bridge the gap in information asymmetry between end users and software systems by mitigating opacity with respect to data practices and thus supporting end users in making more conscious choices.}

The results of this study expanded our understanding of how the individual privacy preferences of users might be retained. \revised{We} were able to derive a set of high-level requirements for privacy explanations. These can serve as a starting point for software engineers to incorporate privacy explanations in software systems. \revised{We conclude that the integration of privacy explanations needs to be conducted carefully to meet the different users' requirements on privacy explanations.}

Furthermore, we propose the trialog of privacy approach as a paradigm for the development of privacy-aware systems since more interdisciplinary collaboration is needed in order to address the complex challenges that arise in terms of privacy.

\section*{Acknowledgments} 
\label{sec:ack}
\addcontentsline{toc}{section}{Acknowledgments}
This work was supported by the Deutsche Forschungsgemeinschaft (DFG, German Research Foundation) under Germany’s Excellence Strategy within the Cluster of Excellence PhoenixD (EXC 2122, Project ID 390833453) and by the research initiative Mobilise between the Technical University of Braunschweig and Leibniz University Hannover, funded by the Ministry for Science and Culture of Lower Saxony.

\bibliography{bibliography}

\end{document}